\documentclass[twocolumn,letterpaper]{IEEEAerospaceCLS}  % only supports two-column, letterpaper format

\usepackage{array} % for the acronym list and macro below
\usepackage{subcaption}
\newcolumntype{P}[1]{>{\raggedright\arraybackslash}p{#1}} %left justified fixed width columns
\usepackage{hyperref}
\usepackage{amsmath}
\usepackage{amsfonts}

\usepackage[]{graphicx}    % default in IEEE format
\newcommand{\ignore}[1]{}  % {} empty inside = %% comment, default in IEEE format

\usepackage[dvipsnames]{xcolor}

\begin{document}
\title{The Safe Trusted Autonomy \\for Responsible Space Program}

\author{%
Kerianne L. Hobbs\\ 
Air Force Research Laboratory\\
2241 Avionics Circle\\
Wright--Patterson AFB, OH, 45433\\
kerianne.hobbs@us.af.mil
\and 
Sean Phillips\\
Air Force Research Laboratory\\
3550 Aberdeen Ave SE \\
Kirtland AFB, NM, 87117
\and 
Michelle Simon\\
Air Force Research Laboratory\\
3550 Aberdeen Ave SE \\
Kirtland AFB, NM, 87117\\
michelle.simon.1@us.af.mil
\and 
Joseph B. Lyons \\
711 Human Performance Wing \\
2210 8th St \\
Wright--Patterson AFB OH 45433 \\
Joseph.Lyons.6@us.af.mil \\
\and
Jared Culbertson\\ 
Air Force Research Laboratory\\
2241 Avionics Circle\\
Wright--Patterson AFB, OH, 45433\\
jared.culbertson@us.af.mil
\and 
Hamilton Scott Clouse\\ 
Air Force Research Laboratory\\
2241 Avionics Circle\\
Wright--Patterson AFB, OH, 45433\\
hamilton.clouse.1@us.af.mil
\and 
Nathaniel Hamilton\\
Parallax Advanced Research\\
4035 Colonel Glenn Hwy\\
Beavercreek, OH 45431\\
nathaniel.hamilton@parallaxresearch.org\\
% Air Force Research Laboratory\\
% 2241 Avionics Circle\\
% Wright--Patterson AFB, OH, 45433\\
%email
\and
Kyle Dunlap\\
Air Force Research Laboratory\\
2241 Avionics Circle\\
Wright--Patterson AFB, OH, 45433\\
kyle.dunlap.5@us.af.mil
\and
Zachary S. Lippay\\
Verus Research\\
45 Hotel Cir NE \\
Albuquerque, NM 87123\\
zachary.lippay@verusresearch.net
\and
Joshua Aurand\\
Verus Research\\
45 Hotel Cir NE \\
Albuquerque, NM 87123\\
joshua.aurand@verusresearch.net
\and
Zachary I. Bell \\
Air Force Research Laboratory\\
203 W Eglin Blvd  \\
Eglin AFB, FL, 32542\\
zachary.bell.10@us.af.mil\\
\and
Taleri Hammack\\
DCS Corporation\\
4027 Colonel Glenn Hwy,  \\
Dayton, OH 45431\\
thammack@dcscorp.com
\and
Dorothy Ayres\\
DCS Corporation\\
4027 Colonel Glenn Hwy  \\
Dayton, OH 45431\\
dayres@dcscorp.com
\and
Rizza Lim\\
DCS Corporation\\
4027 Colonel Glenn Hwy  \\
Dayton, OH 45431\\
rlim@dcscorp.com
\thanks{{U.S. Government work not protected by U.S. copyright.}}%C2022: The MITRE 
}

\maketitle

\thispagestyle{plain}
\pagestyle{plain}

\maketitle

\thispagestyle{plain}
\pagestyle{plain}

\begin{abstract}
The Safe Trusted Autonomy for Responsible Space (STARS) program aims to advance autonomy technologies for space by leveraging machine learning technologies while mitigating barriers to trust, such as uncertainty, opaqueness, brittleness, and inflexibility. This paper presents the achievements and lessons learned from the STARS program in integrating reinforcement learning-based multi-satellite control, run time assurance approaches, and flexible human-autonomy teaming interfaces, into a new integrated testing environment for collaborative autonomous satellite systems. The primary results describe analysis of the reinforcement learning multi-satellite control and run time assurance algorithms. These algorithms are integrated into a prototype human-autonomy interface using best practices from human-autonomy trust literature, however detailed analysis of the effectiveness is left to future work. References are provided with additional detailed results of individual experiments. 
\end{abstract}

\tableofcontents
\begin{figure}
    \centering
    \includegraphics[width=\linewidth]{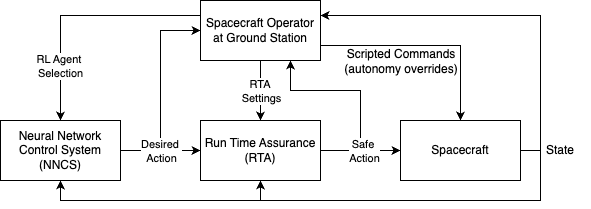}
    \caption{Human operator on-the-loop of a supervised autonomous satellite control concept.}
    \label{fig:ControlLoop}
\end{figure}
%%%%%%%%%%%%%%%%%%%%%%%%%%%%%%%%%%%%%%
\section{Introduction}
%%%%%%%%%%%%%%%%%%%%%%%%%%%%%%%%%%%%%%
Introduction of autonomy technologies in space control has the potential to reduce cost of traditionally human-intensive operations, increase safety by incorporating layered safety assurance control architectures, improve efficiency in operations that do not require a human input, enable a small number of human operators to supervise a larger number of space assets, and increase scientific discovery the pace of scientific discovery. The Safe Trusted Autonomy for Responsible Space (STARS) program seeks to advance autonomy by leveraging state-of-the-art machine learning technologies for decision making and control. However, these tools encounter barriers to trust in the space community, due to uncertainties, opaqueness, brittleness, and inflexibility (including lack of directability). To unleash the full disruptive capability of autonomy, safety, and human-AI teaming, considerations must be incorporated to gain appropriate trust.

The Air Force Research Laboratory’s STARS program is a 3-year seedling for disruptive capabilities program to rapidly advance autonomy technologies for satellite servicing. STARS has the following technical objectives: (1) Develop neural network multi-satellite control and decision-making algorithms using reinforcement learning (RL); (2) Develop run time assurance (RTA) approaches that mitigate hazards and allow the autonomy to stay on the mission; (3) Develop flexible human-autonomy teaming via intuitive and directable interfaces; (4) Develop an integrated testing environment for collaborative autonomous satellite systems and use it to evaluate the first three objectives. The first three objectives impose constraints on each other and are developed concurrently with multiple integration events. Results of the analysis of the RL and RTA designs are summarized with references to publications describing more detailed results. While the prototype interface has been designed using best practices from trust in automation research, analysis of interface design effectiveness is left to future work.  %Each of these four objectives imposes constraints on the other objectives and there are tradeoffs. 

The context for the STARS program is human operator on-the-loop (vs. in-the-loop) control of an autonomous satellite conducting autonomous operations as depicted in Figure~\ref{fig:ControlLoop}. The definition of \textit{autonomous system} is a system that achieves goals within \textit{delegated} and \textit{bounded} authority while operating independently of external control.  \textit{Autonomy} on board the satellite is achieved by a neural network control system (NNCS) trained using RL, which outputs a desired action for the satellite to take based on the current satellite state and mission objectives. \textit{Bounds} are enforced by the RTA, which acts as an automated constraint checker that identifies whether the desired action is within the safety bounds for the mission and either: a) passes through the safe desired action, or b) modifies the action to ensure safety before passing the desired action onto the spacecraft subsystems to implement.  \textit{Delegation} is achieved via a Human-AI teaming Interface (HAI), where the human operator supervising the mission receives information about the spacecraft state, recent desired autonomous system actions, and the actual safe action taken, and interacts with the autonomous satellite by selecting a specific RL agent (i.e. NNCS) trained on the group prior to the mission to have the desired performance for the mission. \textit{Bounds} on the autonomy are specified via the HAI by enabling the human operator to specify safety limits and mission constraints within the RTA component. In the event the performance of the RTA-bounded autonomous agent is unacceptable, the human supervisor may send scripted commands to the spacecraft that override the autonomous control. 

This paper describes the overall STARS program, applicable research from predecessor programs that STARS built directly on, and summarizes the achievements and lessons learned throughout the program for each of the four technical objectives. Detailed results from individual experiments are referenced throughout, and initial HAI designs are presented, while operator studies are left for future work.  First, the exploration of multiple problem formulations and the application and refinement of multiple RL algorithms are presented. Second, a comprehensive RTA approach to simultaneously assure 14 different safety constraints on spacecraft motion, power, thermal, and relationship to other objects in the environment is discussed. A combination of constraints formulated as control barrier functions enforced by an active set invariance filter with a switching RTA for constraints that cannot be formulated in barrier functions is used to enforce constraints (e.g. fuel use constraints).  Third, the iterative design of HAIs via interactions with USSF guardians and the engineering team is discussed. Lastly, autonomous systems need to be verified and tested in relevant laboratory environments. The STARS program led the development of a multi-satellite testing facility to deploy the software and algorithms called the Local Intelligent Network of Collaborative Satellites (LINCS) Lab. The LINCS lab uses aerial drones to emulate a gravity offset for a multi-satellite test and imposes a local space dynamic simulation for trajectories. The laboratory leverages the open-source Robotics Operating System-2 (ROS-2) and Docker services to facilitate integration of the RL, RTA, and HAI technologies above.  Additionally timing tests of the RL-RTA solution on spacecraft processors at the Spacecraft Processing Architectures and Computing Environment Research (SPACER) are discussed.

This paper is organized as follows. First, a historical overview of previous in-space autonomy programs is presented. Second, several RL approaches for a hypothetical inspection mission are presented. Third, development of an RTA system for proximity operations is discussed. Fourth, development of the human-artificial intelligence teaming strategy and interface is described. Fifth LINCS lab development is featured. Sixth, integration and testing of the program components is presented. Finally conclusions are presented.

%%%%%%%%%%%%%%%%%%%%%%%%%%%%%%%%%%%%%%
\section{Background: In-Space Autonomy}
%%%%%%%%%%%%%%%%%%%%%%%%%%%%%%%%%%%%%%
The STARS program was developed for a hypothetical mission and evaluated in simulation. This section puts STARS in the context of related space autonomy programs. It is unique in its proposed use of neural networks for vehicle motion control, novel run time assurance algorithms, and an integrated development approach with human-machine interfaces that prototype human-on-the-loop interactions with neural network-based satellite autonomy. Several previous autonomy technologies, particularly for navigation, proximity operations, and a variety of aspects of Mars rover missions have been tested in space.

The Deep Space One mission, launched by NASA in 1998, was arguably the first use of AI in space \cite{rayman2000results}, where spacecraft health was autonomously monitored, an autonomous navigation system conducted autonomous position and velocity determinations and executed maneuvers to reach a target location, and the Remote Agent Experiment controlled the spacecraft without human supervision for a combined 24 hours. 

There have been several demonstrations of in-space autonomous navigation. Autonomous navigation (AutoNav), which performs image processing, trajectory determination, and maneuver computation, has been conducted on-board spacecraft in many comet and asteroid flyby missions, including Borrelly, Wild 2, Tempel 1, and Hartley 2, and Annefrank \cite{bhaskaran2012autonomous}, as well as to control the trajectory of the Deep Impact mission 370 kg Impactor into the Tempel 1 comet \cite{kubitschek2006deep}. In 2018-2019, the Arcsecond Space Telescope Enabling Research in Astrophysics (ASTERIA) CubeSat conducted three autonomy demonstrations: the use of a task network instead of time-based command sequences, on-board orbit determination without the Global Positioning System (GPS) via passive imaging, and model-based reasoning for in situ hardware health state estimation \cite{fesq2021results}. In 2022, NASA's Artemis I mission successfully completed a demonstration of an autonomous optical navigation system, designed to be a backup navigation system in the event communication with Earth is lost \cite{inman2024artemis}.

Autonomous proximity operations is a challenging area. In 2005, the Demonstration of Autonomous Rendezvous Technology (DART) mission was unsuccessful in demonstrating that pre-programmed and unaided spacecraft could independently rendezvous with a cooperative, passive satellite, and due to a series of faults, unintentionally collided with the target vehicle \cite{croomes2006overview}. In 2007, the Orbital Express experiment demonstrated ``short range and long range autonomous rendezvous, capture and berthing, on-orbit electronics upgrades, on-orbit refueling, and autonomous fly-around visual inspection using a demonstration client satellite" \cite{friend2008orbital}. %\todo{Possible to add: RSGS, NASA OSAM, ANGELS, EAGLE/Mycroft. - got a no from PA/asked to remove ANGELS and EAGLE/Mycroft; OSAM 1 was cancelled, RSGS is behind}

Autonomy has been used extensively in the entry, descent, and landing phases, and for planning and conducting the science missions of Mars rovers. In 2004, the Descent Image Motion Estimation System (DIMES) used terrain-reference navigation to land the Spirit and Opportunity Mars Exploration Rovers (MERs) \cite{johnson2007design}. In 2021, Terrain Relative Navigation (TRN), was used to pinpoint landing of the Mars Perseverance Rover within 5 meters of the targeted location in the Jezero crater \cite{johnson2022mars}. The Autonomous Exploration for Gathering Increased Science (AEGIS) system, first used on the Mars Exploration Rover (MER) Opportunity in 2009 (six years into rover operations), autonomously collected data based on scientist-specified objectives. AEGIS was later used on the Curiosity Mars rover to select targets for the ChemCam remote geochemical spectrometer instrument \cite{francis2017aegis}, and on the latest Mars rover Perseverance to select geochemical and mineralogical analysis targets for the SuperCam instrument \cite{verma2023autonomous}.

%%%%%%%%%%%%%%%%%%%%%%%%%%%%%%%%%%%%%%
\section{Reinforcement Learning}
\begin{figure}[htb!]
    \centering
    \includegraphics[width=0.8\linewidth]{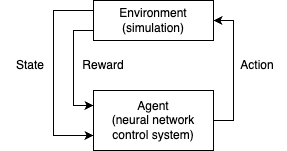}
    \caption{Reinforcement learning feedback loop.}
    \label{fig:RLloop}
\end{figure}

\begin{figure}
    \centering
    \includegraphics[width=0.9\linewidth]{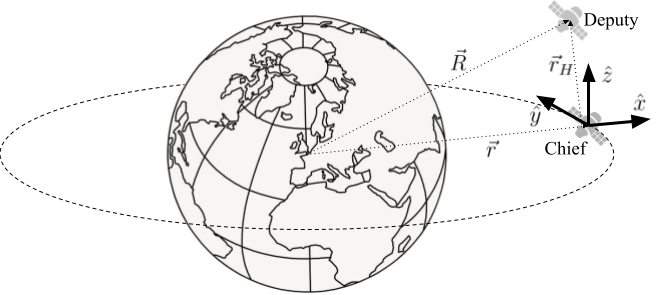}
    \caption{Relative motion Hill's reference frame in which the motion of deputy satellites are described with respect to a chief. The origin of the Hill's frame in centered on the chief spacecraft, $\hat{x}$ is aligned with the vector $\vec{r}$ pointing from the center of the Earth through the chief satellite, $\hat{y}$ is in the velocity direction of the chief satellite in its orbit, and $\hat{z}$ is aligned with the angular momentum vector of the chief in orbit about the Earth.}
    \label{fig:hillsframe}
\end{figure}

%%%%%%%%%%%%%%%%%%%%%%%%%%%%%%%%%%%%%%
RL is a form of machine learning in which an agent (i.e. a NNCS) learns to take actions in an environment based on the state of itself and the environment or observations of the state, and reward for performing as desired, as depicted in Figure~\ref{fig:RLloop}. In STARS, the environment is a satellite proximity operations simulation in which vehicles move according to rigid body spacecraft attitude dynamics \cite{markley2014fundamentals} and linearized Clohessy--Wiltshire dynamics \cite{clohessy1960terminal} in Hill's reference frame\cite{hill1878researches}, depicted in Figure~\ref{fig:hillsframe}. The agent, a NNCS, takes actions in the form of thrusting or rotating in this environment, receiving rewards to achieve the desired goal. After some number of episodes, the RL algorithm updates the neural network state-action pairing to improve rewards earned. This update occurs repeatedly throughout the training. 

\subsection{Environment Development and Initial Reinforcement Learning Solutions}

An incremental approach was taken to the formulation of satellite proximity operation challenges into environments, conforming to the OpenAI Gym API standard \cite{brockman2016openai}, as well as developing RL solutions to those environments. 

\subsubsection{Docking Challenge Problems} Initially the team developed a series of docking proximity operations RL challenge problems, inspired by related autonomous rendezvous and proximity operations challenge problems developed for hybrid control and estimation approaches \cite{jewison2016spacecraft}. First, a challenge problem was presented to develop a controller to maneuver an underactuated spacecraft to the origin under linearized, planar CWH equations using thrust in the body x-direction only and a single reaction wheel \cite{petersen2021challenge}. Next, several safe RL benchmarks \cite{ravaioli2022safe} were presented describing the dynamics, metrics, a distant-dependent speed limit safety constraint, implementation in an OpenAI Gym API, and initial RL results for three docking tasks: docking in planar (2D) CWH equations, docking in 3D under CWH, and the underactuated satellite presented in the earlier challenge problem. A variety of solutions were developed for the docking challenge problems, including initial solutions to the benchmarks \cite{ravaioli2022safe} and a comparison of RL to fuzzy control for the docking problem \cite{dunlap2022comparing}.
In \cite{chenRL}, we partnered with Dr. David Copp at the University of California Irvine to leverage a Proximal Policy Optimization (PPO) RL scheme for a 3D trajectory planning for docking. We consider constraints on the maximum thrust and line-of-sight constraints and an actor/critic methodology for training. In \cite{chenRL}, we thoroughly analyze and present statistics on performance like mission time, fuel consumption, and computational time. 

\subsubsection{Inspection Challenge Problems} After a series of docking challenge problems, the team moved to an inspection challenge problem in which one or more deputy spacecraft needed to maneuver around a chief to complete an inspection. For these problems, it was assumed that a point on the spacecraft was ``inspected" if it was within line of sight for the deputy spacecraft and illuminated by the sun. First, an illumination model was created and a PPO \cite{schulman2017proximal} RL algorithm was used to train an NNCS to maneuver a simulated deputy spacecraft using only translational motion control for autonomous inspection \cite{vanWijkAAS_23}. Ultimately, a six degree-of-freedom environment was generated that combined the translational and rotational dynamics and trained an RL agent to complete the inspection task \cite{dunlap2024run}, and it was later extended to the multiagent scenario with varying numbers of controllable deputies \cite{dunlap2025multiagent}. Additional variations of the inspection environment are discussed within this section.

\subsection{Ablation Studies}
In addition to developing RL environments and training agents using existing RL algorithms, the STARS team sought to better understand how RTA during training and different rewards, observation space, and action space formulations impacted the performance of trained agents. First, the team sought to understand how best to incorporate RTA into the training of the RL agents by studying how different configurations and reward structures impacted the overall performance \cite{hamilton2023ablation}. The results showed RTA can be useful for guiding agents to safe and successful behavior, however if the behavior is not reinforced through reward function (e.g. assigning a cost to RTA intervention), agents will become dependent on the RTA to take over in riskier states instead of executing the safe behavior on their own. Second, the team studied the impact of having the RL agent choose from a set of discrete thrust values instead of selecting any continuous thrust value between a defined minimum and maximum \cite{hamilton2024investigating}. The results highlighted how the learning task can dictate which action space is more advantageous. For the inspection task, where optimal behavior uses sparse adjustments to the orbit, the discrete action space is the best fit. In contrast, the docking task requires frequent, fine-tuned adjustments during the approach, working best with the continuous action space. Third, the team wanted to understand how differences to the observation space, through reference frame and sensor selection, impacted the training \cite{hamilton2024obs}. The results showed how versatile RL is at learning to inspect the majority of inspection points even with limited information. The additional sensors were not necessary for learning to inspect all points, but helped produce more fuel efficient policies and improved sample complexity (i.e. agents learned quicker). Additionally, the results showed that the consistency in reference frames is more important than which is selected, but Hill's reference frame led to the best behaviors.
\begin{figure*}[htb!]
\centering
\includegraphics[width=.8\linewidth]{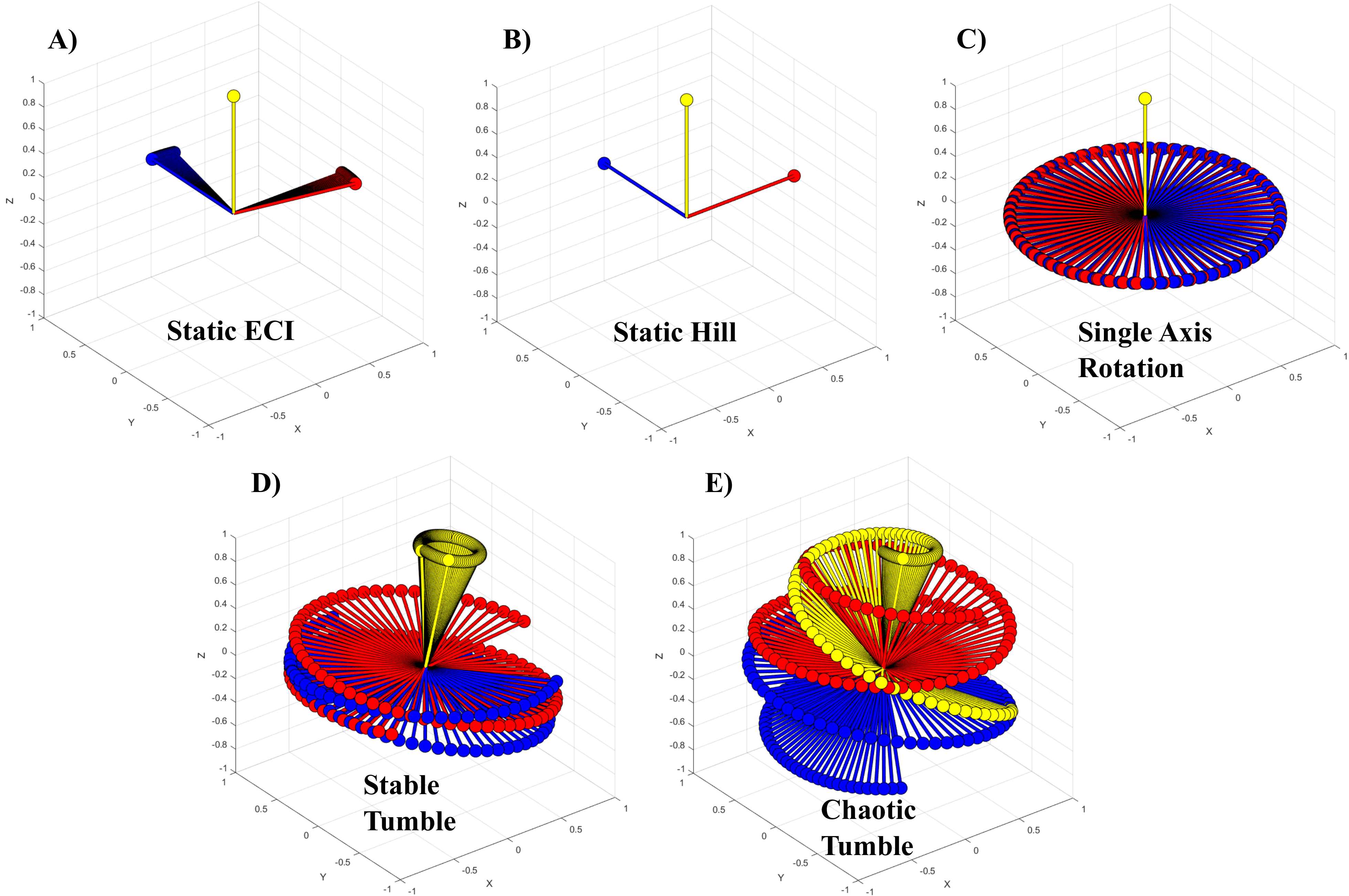}
\caption{Examples of torque-free rigid body dynamic behaviors. Shown are the body fixed axes as viewed in Hill's frame. Each example was generated over a different amount of time to exemplify distinctive qualitative features of each mode.}
\label{fig:CombinedDynamics}
\end{figure*}

\subsection{Hierarchical Reinforcement Learning}
There are some notable tradeoffs when exploring the relationship between low-level actuator learning and high-level decision-making for space exploration. The STARS team explores these tradeoffs more closely in \cite{Aurand24,10.1007/978-3-031-51928-4_76,aurand2023exposurebasedmultiagentinspectiontumbling}. Within these works, \textit{hierarchical} RL is explored as a means of separating low-level actuation from high-level decision-making. Therein, high-level decisions are made through waypoint-based guidance generation, reflecting the maximization of joint information retrieval. Low-level actuation is then responsible for moving each inspecting agent between subsequent waypoint targets. In \cite{10.1007/978-3-031-51928-4_76}, the case of three homogeneous inspector satellites was considered. The agents were trained by building a multi-agent Markov Decision Process (MDP) to traverse 20 fixed inspection points. In \cite{10.1007/978-3-031-51928-4_76}, a trade off was found between expected trajectory arclength and the ability for agents to effectively learn collision avoidance. When straight-line distance between two waypoints is small ($<$250m) and there is a high probability of vehicle intersection with another agent, a collision event is effectively guaranteed. With this in mind, the team began investigating cases where the inspected agent was tumbling, described under relative attitude coordinates. Tumbling itself can be characterized in several ways, from a single-axis stabilized rotation to a chaotic tumble. Figure~\ref{fig:CombinedDynamics} shows the evolution of multiple types of relative rotations. In \cite{aurand2023exposurebasedmultiagentinspectiontumbling}, the team trained the hierarchical policy to operate under a variety of tumbling scenarios and provided a detailed analysis of the trade-off between inspection time and $\delta V$ used. In \cite{Aurand24}, this was expanded to do an exhaustive hyper parameter search for each tumbling mode outlined in Figure~\ref{fig:CombinedDynamics} (Static CHW, Static ECI, SingleAxis Rotation, Stable Tumble, Chaotic Tumble).

\subsection{Dynamic Multi-objective Reinforcement Learning}
A critical component that STARS addressed is the multi-faceted nature of the objectives in the satellite inspection problem. For example, efficient fuel management is essential in space operations, but must be balanced against completion of the inspection mission and safety. Moreover, the prioritization of these objectives might be dynamic within a single mission, requiring a flexible approach to training that allows an operator to choose at run-time a relative weighting of mission goals. In \cite{susfas}, the STARS team studied the viability of a stacked variation of a universal successor features approximation approach (based on existing work \cite{uvfa}) coupled with the STARS-developed secondary RTA controller from \cite{dunlap2023RTA_inspection} which allows training over a range of weightings on the components of a reward function. In addition to being able to direct agent behavior to match operator weight preferences, this study found that the resulting trained agents were able to achieve significant improvements in fuel usage over baseline RL agents, while maintaining high inspection performance.

\subsection{Real-Time Improvement for Reinforcement Learning}
The problems solved by the various model-free reinforcement learning methods and architectures has shown the benefits of using reinforcement learning; however, existing reinforcement learning solutions rely on offline training and may fail if the state or dynamics are outside the domains they're trained. RTA controllers are one method to ensure system stability when unexpected circumstances cause changes to the system. An additional focus of the STARS program has been to develop stable, real-time adaptive reinforcement learning methods that monitor a system and adapt trained policies for changes in dynamics. The work in \cite{makumiBellCSS2023} examined one technique for estimating unknown interaction dynamics in real time between a herding agent and evading agent and approximating the optimal policy for herding the evading agent to a desired goal state using online reinforcement learning methods. The result showed significant improvement over a baseline method.

\subsection{Neural Network Verification Efforts}
Another path that the STARS team investigated was whether it might be possible to directly verify properties of the NNCS's behavior rather than relying on the RTA to assure safety. In the formal methods community defines properties in categories of liveness (something good will eventually happen) and safety (something bad will never happen). The STARS team partnered with researchers at Stanford to apply formal analysis techniques to a 2D docking neural network to prove the liveness property that the deputy under control of the neural network would eventually dock with the chief spacecraft, no matter where it started within a bounded region around the chief and to prove that the deputy would never crash (dock at speed in excess of a safe value) with the chief \cite{mandal2024safe}. This neural network verification approach is a state-of-the-art achievement, but also highlights the limits of the current approaches to systems with small observation and action spaces, as well as relatively simple dynamics.

% \todo{Sean/Josh - this section is for you\cite{Aurand24,10.1007/978-3-031-51928-4_76} Deep Q learning for decentralized multi-agent inspection of a tumbling target}
% \todo{Exposure-Based Multi-Agent Inspection \cite{aurand2023exposurebasedmultiagentinspectiontumbling}}

%%%%%%%%%%%%%%%%%%%%%%%%%%%%%%%%%%%%%%
\section{Satellite Platform Safety and Run Time Assurance}\label{sec:RTA}
%%%%%%%%%%%%%%%%%%%%%%%%%%%%%%%%%%%%%%

\begin{figure}[htb]
    \centering
    \includegraphics[width=\linewidth]{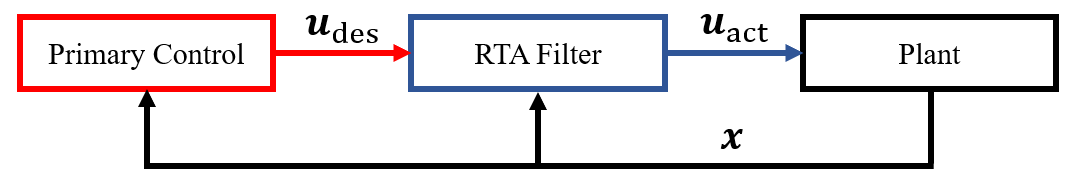}
    \caption{Run time assurance control system architecture}
    \label{fig:RTA_Filter}
\end{figure}

RTA is an online safety assurance technique that filters an unverified primary controller to ensure system safety \cite{hobbs2023runtime}. The RTA component is agnostic to the primary controller, which can take a wide variety of forms, including but not limited to: human operators, traditional control theory designs, adaptive control systems, or AI controllers. For STARS, neural network primary controllers are developed using RL \cite{dunlap2022run}. RTA is particularly useful in situations where the primary controller is too complex to completely verify offline at design time.  
%This allows the control system to separate the objectives of performance and safety assurance.
As depicted in Figure~\ref{fig:RTA_Filter}, RTA acts inside the control loop by monitoring the state $\boldsymbol{x}$ of the plant (in this case the states of the vehicles in the mission) and the desired control signal $\boldsymbol{u}_{\rm des}$ from the primary controller. The RTA filter outputs an actual control signal $\boldsymbol{u}_{\rm act}$ that is either a) the unchanged desired control $\boldsymbol{u}_{\rm des}$ if it meets safety constraints, or b) a modified control command that ensures safety when necessary. A rudimentary form of RTA currently used in spacecraft is ``safe mode" or ``sun safe mode," which attempts to reestablish ground contact while charging solar panels when the spacecraft experiences a fault. In STARS, the team developed a more nuanced set of interventions and backups to keep the spacecraft on mission while adhering to safety constraints in the event the primary control sends an unsafe command.
%and modifies the control if necessary.  RTA Inside the control loop, the RTA filter analyzes the output of the primary controller based on the state and modifies it as necessary to ensure it is safe.In this figure, the control inputs $\boldsymbol{u}_{\rm des}$ and $\boldsymbol{u}_{\rm b}$ are defined based on design tools, however, in this scheme, there is a logic state that takes into account the current state, safety requirements, and, potentially, future propagated states. The job of the RTA is to switch between a verified backup controller and a primary controller.

%\setlength{\unitlength}{1\linewidth}
%\begin{figure}
%    \centering
%    \includegraphics[width=\unitlength,trim = 10mm 30mm 10mm 10mm, clip]{Figures/Tech/RTA-general.png}
%    \put(-.37,.3){\scalebox{.6}{$\boldsymbol{u}_{\rm act}(x)$}}
%    \put(-.7,.3){\scalebox{.6}{$\boldsymbol{u}_{\rm des}(x)$}}
%    \put(-.49,.19){\scalebox{.6}{$u_{b}(x)$}}
%    \caption{Run time assurance control system architecture.}
%    \label{fig:RTA}
%\end{figure}

For a continuous-time control affine dynamical system modeled as,
\begin{equation} \label{eq:fxgu}
   \dot{\boldsymbol{x}} = f(\boldsymbol{x}) + g(\boldsymbol{x})\boldsymbol{u}
\end{equation}
safety can be defined in terms of a set of $M$ inequality constraints, $h_i(\boldsymbol{x}): \mathcal{X} \to \mathbb{R}$,  $\forall i \in \{1,...,M\}$, where $\mathcal{X}$ is the set of all possible state values and $h_i(\boldsymbol{x}) \geq 0$ when a constraint is satisfied. This set of constraints is referred to as the \textit{safe set}, $\mathcal{C}_{\rm S}$, and is defined as,
\begin{equation}
    {\mathcal{C}_{\rm S}} := \{\boldsymbol{x} \in {\mathcal{X}} \, | \, h_i(\boldsymbol{x}) \geq 0, \forall i \in \{1,...,M\} \}
\end{equation}

Two common types of RTA filter intervention mechanisms are \textit{switching}, sometimes also called a Simplex \cite{simplex_1} architecture, and \textit{optimization}, where a popular choice is the Active Set Invariance Filter (ASIF) \cite{gurriet2018online,dunlap2022comparing}. Switching RTAs output the primary controller $\boldsymbol{u}_{\rm des}$, as long as the flow $\phi_1^{\boldsymbol{u}_{\rm des}}(\boldsymbol{x})$ (e.g. trajectory) of the system under that controller will stay in the safe set $\mathcal{C}_{\rm S}$, otherwise the system will switch to a verified backup controller $\boldsymbol{u}_{\rm b}$ to assure safety of the system. This is formally described in Equation \ref{eq:switching}. On the other hand, optimization filters minimize the deviation of the actual control $\boldsymbol{u}_{\rm act}$ from the desired control $\boldsymbol{u}_{\rm des}$ using a Quadratic Program (QP) subject to barrier constraints $BC_i$ on the state and control for $M$ safety constraints, as formally described in Equation \ref{eq:optimization}. The barrier constraints are defined with using Control Barrier Functions (CBFs) \cite{ames2019control},
\begin{equation}
    BC_i := \sup_{\boldsymbol{u}\in \mathcal{U}}[L_f h_i(\boldsymbol{x}) + L_g h_i(\boldsymbol{x})\boldsymbol{u}]  +\alpha(h_i(\boldsymbol{x})) \geq 0
\end{equation}
where $\mathcal{U}$ is the admissible control set, $L_f$ and $L_g$ are Lie derivatives of $h_i$ along $f$ and $g$ respectively, and $\alpha(x)$ is a continuous, strictly increasing class $\kappa$ function. The objective of the barrier constraint is to analyze the boundary of the safe set to ensure that $\dot{h}_i(\boldsymbol{x}) \geq 0$, and therefore ensures that $\boldsymbol{x}$ will never leave $\mathcal{C}_{\rm S}$.

\begin{samepage}
\noindent \rule{1\columnwidth}{0.7pt}
\noindent \textbf{Switching Filter}
\begin{equation}
\begin{array}{rl}
\boldsymbol{u}_{\rm act}(\boldsymbol{x})=
\begin{cases} 
% \boldsymbol{u}_{\rm des}(\boldsymbol{x}) & {\rm if}\quad \boldsymbol{x} \in \mathcal{C}_{\rm S}  \\
\boldsymbol{u}_{\rm des}(\boldsymbol{x}) & {\rm if}\quad \phi_1^{\boldsymbol{u}_{\rm des}}(\boldsymbol{x}) \in \mathcal{C}_{\rm S}  \\ 
\boldsymbol{u}_{\rm b}(\boldsymbol{x})  & {\rm if}\quad otherwise
\end{cases}
\end{array}\label{eq:switching}
\end{equation}
\noindent \rule[7pt]{1\columnwidth}{0.7pt}
\end{samepage}

\begin{samepage}
\noindent \rule{1\columnwidth}{0.7pt}
\noindent \textbf{Optimization Filter}
\begin{equation}
\begin{split}
\boldsymbol{u}_{\text{act}}(\boldsymbol{x})=\underset{\boldsymbol{u} \in \mathcal{U}}{\text{argmin}} & \left\Vert \boldsymbol{u}_{\text{des}}-\boldsymbol{u}\right\Vert_2 ^{2}\\
\text{s.t.} \quad & BC_i(\boldsymbol{x},\boldsymbol{u})\geq 0, \quad \forall i \in \{1,...,M\}
\end{split}\label{eq:optimization}
\end{equation}
\noindent \rule[7pt]{1\columnwidth}{0.7pt}
\end{samepage}

Additionally, safety monitoring can be defined \textit{explicitly} or \textit{implicitly}. Explicit constraints are developed offline using a precise, mathematically defined, safe set. Implicit constraints are developed online using trajectories under a backup control law. A comparison of these RTA approaches for spacecraft docking with a classical control primary controller is discussed in \cite{dunlap2021comparing} and with an RL primary controller in \cite{dunlap2023run}.

Software implementations of RTA are difficult to develop by hand and often suffer from scalability and portability challenges, especially advanced ASIF techniques that use methods such as implicit CBFs and high order CBFs. To combat these challenges, the STARS team created the Safe Autonomy Run Time Assurance Framework \cite{ravaioli2023universal} to easily build and deploy RTA systems in Python. The framework uses JAX \cite{jax2018github} to automatically compute derivatives required for ASIF methods, which minimizes user effort and error.

\subsection{Safety for Rendezvous and Proximity Operations}
Safety in RPO is a complex operation. There are unique safety considerations that must be considered to ensure that the satellite platforms are safe. In \cite{10155826}, the authors considered safety from the perspective of guidance, navigation, and control partitioning wherein, each component was separated and relevant safety measures were applied. The article was a brief tutorial paper at the 2023 American Control Conference and outlined several state-of-the-art methods that encompass implicit or explicit safety within the design. 

The safety constraints used in STARS were initially elicited from a risk-based hazard analysis, with rationale in \cite{hobbs2020elicitation,hobbs2021risk}. While the initial intent of these requirements generally stayed intact throughout the program, the formal interpretation of the meanings varied depending on the RTA implementations and varying use cases. 
% \todo{overview of safety constraints}

While many safety constraints could be considered, the following were developed for different space tasks.
\begin{enumerate}
    \item \textit{Safe separation}: The deputy shall not collide with the chief or another deputy.
    \item \textit{Dynamic speed constraint}: The deputy shall decrease its speed as it approaches the chief. This minimizes risk in the event of a fault.
    \item \textit{Keep in zone}: The deputy shall not travel too far from the chief.
    \item \textit{Passively safe maneuvers}: In the event of a fault or loss of power, where the deputy cannot use its thrusters, the deputy's natural motion shall not cause a collision with the chief over an extended time period.
    \item \textit{Axial velocity limits}: The deputy shall not maneuver aggressively with high velocities.
    \item \textit{Attitude exclusion zone}: The deputy shall not point sensitive sensors at the Sun, to avoid instrument blinding.
    \item \textit{Communication}: The deputy shall maintain attitude requirements for communication with a ground station.
    \item \textit{Maintain acceptable temperature}: The deputy shall ensure all components do not become too hot or too cold.
    \item \textit{Maintain battery charge}: the deputy shall maintain attitude requirements for sufficient power generation.
    \item \textit{Angular velocity limits}: The deputy shall not maneuver aggressively with high angular velocities.
    \item \textit{Fuel limit}: The deputy shall adhere to a maximum cumulative fuel use limit.
\end{enumerate}

\subsection{Run Time Assurance for Proximity Operations}
An initial RTA architecture was developed with a set of mode logic, and the constraints were formulated in past time linear temporal logic for verification \cite{hobbs2020elicitation,hobbs2021formal}.

For multiagent inspection, the constraints related to translational motion are enforced with ASIF in \cite{hibbard2022guaranteeing} and \cite{dunlap2023RTA_inspection}. For a single agent, the constraints related to attitude are enforced with ASIF in \cite{McQuinn2024RTA}. Additionally, one common orbit that satisfies many of the listed constraints is a closed elliptical Natural Motion Trajectory (eNMT). eNMTs are often considered safe parking orbits, as the deputy can remain in an elliptical orbit around the chief without using any fuel. This ensures the deputy will never collide with the chief. RTA for navigating to eNMTs is developed in \cite{mote2021natural}.

In \cite{10365665}, the team leveraged tools like mixed monotonicity and invariance properties to guarantee that inputs will never cross a safety barrier. In particular, the problem of attitude control was considered to maintain sensor safety and keep out zones regardless of the primary controller input. A backup controller is leveraged to provide appropriate inputs $\boldsymbol{u}_{\rm act}$ into the plant. An RTA scheme is then proposed and sufficient conditions for a class of design choices are proposed. 

A series of projects followed which expanded on existing RTA methods, and utilized them for related applications. Building on the results in \cite{vanWijkAAS_23}, \cite{vanWijk2024JAIS} developed discrete CBFs to ensure safety of an RL-trained neural network tasked with performing single agent inspection. A related paper developed a fault-tolerant RTA scheme for handling abrupt state change faults in the presence of angular rate constraints \cite{vanWijk_FTRTA}. To increase robustness of the RTA under uncertainty from bounded, unmodeled disturbances, a disturbance-robust backup control barrier function method was developed in \cite{vanWijk2025ACC}. This general technique provides guarantees for a wide range of nonlinear systems and constraints. 

Safety like those considered in \cite{10155826} can be implicitly developed in the policies governing satellite behaviors. In the STARS program, there has been significant development and consideration of analytical safety guarantees. In \cite{10156074,miller2024acc,hibbard2023trajectory}, the selection of control policies were considered to provide safe behaviors. More specifically, this work synthesizes the correct policy based on the current state and the objective to mathematically guarantee safety. In \cite{10156074,miller2024acc}, a fully actuated 6 DOF relative motion satellite model was considered for a reach-avoid scenario, wherein, there were safety keepout zones for both attitude and translational maneuvering. In \cite{hibbard2023trajectory}, avoiding the alignment of sensors with the sun was considered, such that they avoid colliding with one another, and that they avoid colliding with the chief. In \cite{hibbard2023trajectory}, the system was constructed using an MDP model to determine an optimal set of waypoints to maneuver for inspection \cite{10365665}.

Computational chipsets on space systems are typically rudimentary compared to their terrestrial counterparts and, typically, for decision-making and scheduling, optimization policies reign supreme to provide those optimal or feasible solutions. However, such tools tend to be computationally burdensome. In \cite{BEHRENDT20239380}, the case of optimization-based policies was considered for RPO where the computational aspect was investigated. In particular, a detailed analysis was provided of the number of computational iterations to maintain feasibility and the tradeoff between that and optimality. It is shown that a near-optimal solution can be maintained by considering only 5-6 iterations of the optimization solver, dramatically increasing computations. Another way to achieve faster compute execution times are the RL policies considered within this paper. Because the NNCSs trained using RL are relatively small, and all the training is done on the ground, their computational overhead is also very small at the edge (on board the satellite). In Section~\ref{sec:labtesting}, some RL and RTA policies on space-rated hardware are investigated.

%%%%%%%%%%%%%%%%%%%%%%%%%%%%%%%%%%%%%%
\section{Human-Artificial Intelligence Teaming and Interface}
%%%%%%%%%%%%%%%%%%%%%%%%%%%%%%%%%%%%%%
The proposed human operator on-the-loop architecture for future missions, which may incorporate increasing autonomy and new technologies such as machine learning, represents a significant paradigm shift in spacecraft operations that have traditionally relied on human-centered planning far in advance. This could create barriers to operator trust which the team sought to address through the combination of directability, bounded autonomy, enhanced situation awareness, and calibrated trust in the autonomous agent's decision making at execution time. The STARS team developed a prototype HAI for operators to team with advanced autonomy algorithms to complete a mission. A key consideration for Human-Machine Teaming (HMT) is that the interactions between the human and machine maintain a degree of flexibility to allow the human to direct and modify the machine’s behaviors whenever needed \cite{calhoun2022adaptable}, and enable responsible use of machine learning-based control \cite{lyons2023responsible}.

Li et al. \cite{li2021individualized} have also examined Human-autonomy teaming in the presence of a machine-learned autonomy, finding that within Human-AI Teams is contingent on ensuring AI agent alignment with human goals. They note a significant challenge remains in designing an AI agent that adapts to human goals that change over time as the context evolves. To evaluate their hypothesis, they tested multiple Human-AI teams in an interdependent game. They found that adaptive policies (i.e., those that were capable of recognizing the human’s role and shifts in that role) were associated with better performance relative to a static (i.e., where the agent would remain in one role) or a random (i.e., where the agent would randomly adopt a new role) agent policy. The current manuscript adopts a similar philosophy, however our approach to ensuring goal alignment was derived through the directability dashboard rather than the agent having to learn the human’s policies.

\subsection{Directability and bounded autonomy}
\textit{Directability} refers to the ability of a human supervisor to define policies that influence agent activities during execution of a mission \cite{myers2001human}. The team recognized that the opportunities for directability were most plentiful for the RTA algorithms rather than the RL algorithms. Initially the team hoped to incorporate real-time weighting of mission goals in the RL solution, as described in the section titled ``Dynamic Multi-objective Reinforcement Learning." However, in practice the maturity of the approach was not high enough yet to incorporate into the interface. Instead, an approach was implemented where an operator may be able to select a pre-trained RL agent that has a particular weighting of objectives (e.g., more fuel conservative versus more aggressive) but there are currently few opportunities to direct specific actions in real-time once a RL agent is selected. In contrast, the RTA algorithms offered multiple opportunities to set and modify priority, to set and modify specific parameters, and to turn certain parameters on or off in the most simple sense. Further, operators could easily toggle priorities among the subset of RTA parameters to seamlessly set a higher or lower priority among the set.

\subsection{Bounded Autonomy}
\begin{figure}
    \centering
    \includegraphics[width=0.7\linewidth]{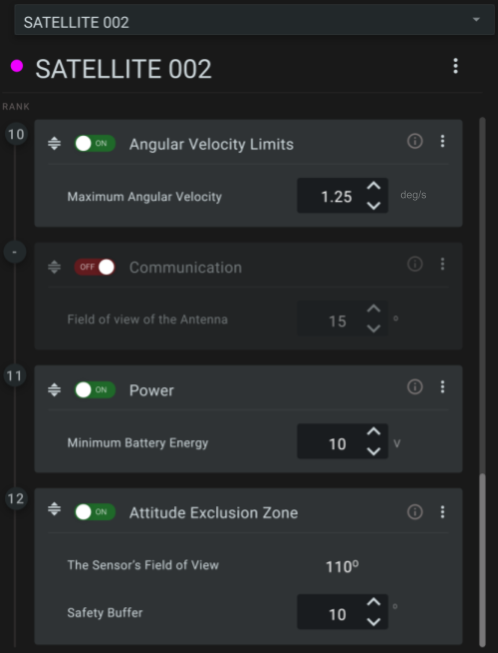}
    \caption{Mockup of the run time assurance settings in the human AI interface, where users can click and drag titles up or down to change the priority of constraints, turn constraints on or off depending on the mission and set specific values for the constraints.}
    \label{fig:ranking}
\end{figure}

\begin{figure}
    \centering
    \includegraphics[width=0.7\linewidth]{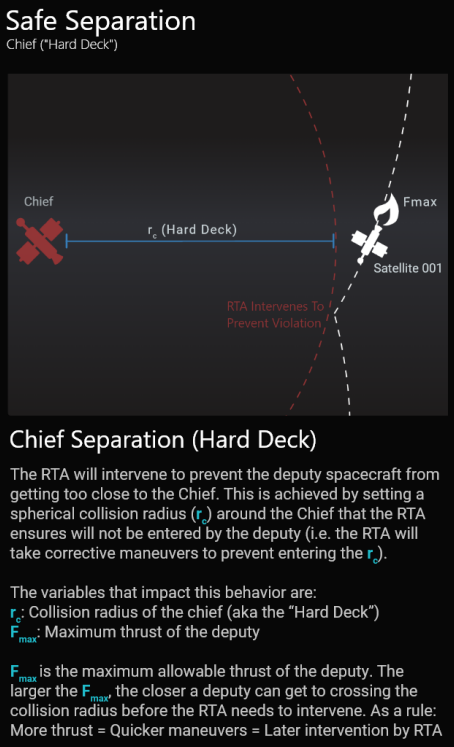}
    \caption{Mock up of a help card depicting the variables used to define a safety constraint in the run time assurance algorithm.}
    \label{fig:helpcard}
\end{figure}

\begin{figure}
    \centering
    \includegraphics[width=0.7\linewidth]{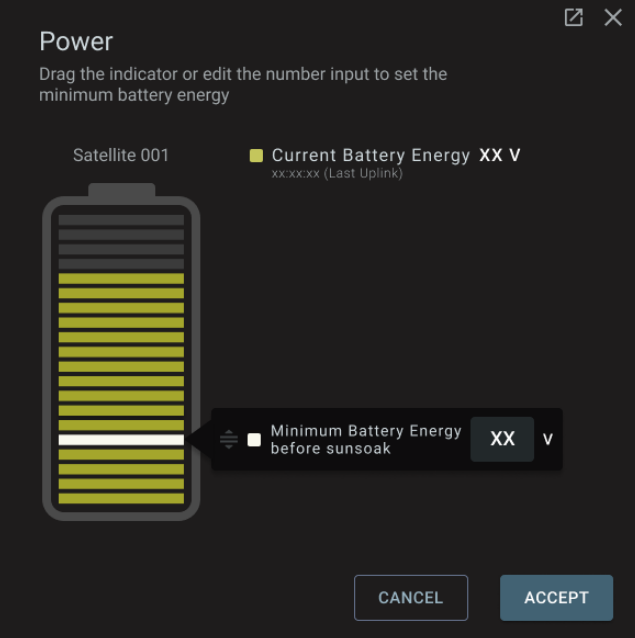}
    \caption{Mock-up of an interactive pictorial depiction of the Power constraint where users can type the minimum power or drag and drop on the figure.}
    \label{fig:power}
\end{figure}

Autonomous decision making is bounded by safety and mission constraints set in the RTA window. Within this window, depicted in Figure~\ref{fig:ranking}, the operator can:
\begin{itemize}
    \item rank order the priority of constraints defined in Section \ref{sec:RTA} by dragging tiles up and down or by clicking and entering a number.
    \item set specific values for constraints
    \item click on the circled ``i" information icon for a help card that describes how that particular constraint is enforced, and gives a pictorial description of what variables are being set, as depicted in Figure~\ref{fig:helpcard}, and
    \item click on the 3 vertical dots to bring up a larger interface with a pictorial depiction of the constraint in context of that resource on the vehicle, as shown in a mockup of the power constraint in Figure~\ref{fig:power}. 
\end{itemize}

\subsection{Situational Awareness and Calibrated Trust}
\begin{figure}
    \centering
    \includegraphics[width=0.9\linewidth]{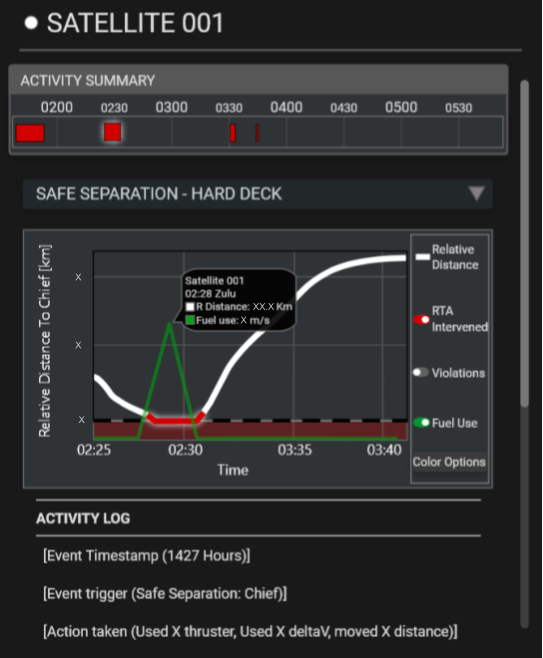}
    \caption{Mock-up of a timeline view that describes when and why RTA activated to assure mission constraints are satisfied. In this example, RTA intervened to ensure that the spacecraft maintained a minimum safe distance from another spacecraft. }
    \label{fig:timeline}
\end{figure}

In the STARS interface, operators may input preferences for a mission such as time efficiency or fuel efficiency and several courses of action are presented to the operator depicting anticipated spacecraft trajectories – enabling predictability of the algorithms.  Each course of action is provided by a different algorithm that offers a more tailored solution based on a variety of training conditions with different priorities in performance and starting states. The operator can then select from one of the provided courses of action or substitute a manually-defined plan. The goals of the interface are also to increase the operator's situational awareness of the algorithms and facilitate calibrated trust. For \textit{situational awareness}, the goal is for the operator to understand what the RTA-bounded neural network system commanded, why it acted that way and what it is likely to do next. These factors are key elements relating to perception, understanding, and projection – which are the tenets of the situation awareness-based model of transparency \cite{chen2018situation}. For \textit{calibrated trust}, the goal of the interface is to provide information that enables operators to trust the autonomy when it is performing well, but also recognize when it is being utilized in scenarios it was not trained on (i.e., when it is outside of its capabilities). Tools like the timeline view depicted in Figure~\ref{fig:timeline} are envisioned to enable operators to know when and why RTA intervened to assure mission constraints are satisfied. Methods such as these may help to facilitate shared mental models of the machine’s capabilities, which should enable calibrated trust from human partners of the agents \cite{lyons2021human}. In the future, agents may need information about their human partners to build computational representations of their human partners as well. Additional research is required to understand whether the prototype HAI promotes situational awareness of the user, facilitates calibrated trusts, and to understand where this combination of autonomy technology lies on a space trusted autonomy readiness scale \cite{hobbs2023space}.

% \todo{Sean/Zack - Add a brief overview of integration}

%%%%%%%%%%%%%%%%%%%%%%%%%%%%%%%%%%%%%%
\section{Laboratory Development, Integration, and Testing}\label{sec:labtesting}
%%%%%%%%%%%%%%%%%%%%%%%%%%%%%%%%%%%%%%
Through this program, the STARS team integrated their developed technologies with different hardware and software instances both at AFRL and in academia. In particular, this section discusses development of the LINCS lab and integration of the RL, RTA, and HMI with the LINCS lab and in-flight hardware through the SPACER Lab at AFRL. Additionally, tests of attitude safety guarantees deployed in the ASTROS Lab at The Georgia Institute of Technology are discussed.

%%%%%%%%%%%%%%%%%%%%%%%%%%%%%%%%%%%%%%
\subsection{Development of the Local Intelligent Network of Collaborative Satellites Laboratory}
%%%%%%%%%%%%%%%%%%%%%%%%%%%%%%%%%%%%%%
Due to the prohibitive cost of launching satellites, even with the existence of launch capabilities and opportunities, many organizations are not willing to launch experimental autonomous algorithms and software to command and control satellite behaviors. That said, there exist numerous terrestrial satellite emulation environments that offset Earth's gravity.  In particular, there exist several terrestrial-based environments for the verification, validation, testing, and evaluation of advanced satellite algorithms. These systems range from pure software solutions to N degree-of-freedom large robotic arm facilities. Since the goal of each platform is to most closely represent the frictionless free-fall environment of the space domain, but within an environment that is inherently not so, each solution has its pros and cons. In \cite{WILDE2019100552}, the authors provide a concise outline of such limitations of satellite testbeds ranging from limitations in degrees of freedom in orientation or translation capabilities, the number of simulated objects, and the mechanics of simulation/emulation. To summarize, there is no perfect scenario for terrestrial satellite testbeds and there are tradeoffs to each configuration. 

At a high-level, these testbed environments can be partitioned into several different categories: 1) simulation or software-in-the-loop (SIL), 2) processor-in-the-loop (PIL), or 3) hardware-in-the-loop (HIL). Several software simulation tools have been developed spanning government, industry, and academia, \cite{stoneking201842,kenneally2020,agi-stk-website,freeflier-website,gmat,9438538,10115695,cameron2016}. A frequently used technique for ensuring that the flight software used onboard the satellite system works as intended is called a `flat sat,’ which is also typically referred to as a PIL. In the PIL, the dynamics, sensors, and actuators are typically simulated using a simulation environment, but the flight software is uploaded onto the processing hardware. Tests are run to determine nominal and peak power draws, the feasibility of the software/hardware combination, and a variety of other metrics \cite{10115695,Martins-Filho14,wilde2016,NRL-SpaceRobot,cho2009,tsiotras2014astros,agrawal2001air,WILDE2019100552,schwartz2003,gaias2010hardware,grubb2016nasa,9109954,doi:10.2514/6.2024-0508}. However, the largest limitation of these simulation platforms is the lack of any physics or realistic sensor data within the loop. A detailed survey of recent close-proximity space robotics emulators is provided in \cite{WILDE2019100552} and highlights many specific satellite test platforms. However, it is worth noting that in \cite{WILDE2019100552}, there is a distinct lack of leveraging the advances in small lightweight aerial vehicles to offset gravity and provide the basis platform for terrestrial testing.

To address the issues of testing collaborative autonomy in a safe environment, the team investigated the validity of leveraging terrestrial Class-I aerial vehicles as a proxy for satellite systems. 
In STARS, the team has built a novel testing capability called the Local Intelligent Network of Collaborative Satellites (LINCS) Lab. 
In this lab, the aerial platforms are forced to behave like satellite systems (i.e., emulate satellites) by forcing them to follow 2-body dynamics with J2 perturbations, and resolve them in a relative frame (i.e., Hill's frame) which can be scaled into the lab setting. 
%The Hill frame is a moving relative reference frame where the state represents a position with respect to a chief (sometimes referred to as a target) satellite. (this is described in the RL section already)
Once these dynamics are applied to the agents, close proximity algorithms can be applied to the emulated satellite systems without regard to the aerial vehicles, thus, providing insight into how these algorithms would act on orbit. The laboratory has been extensively discussed in \cite{doi:10.2514/6.2024-1207} and provides initial numerical results therein. 

\begin{figure*}
    \begin{subfigure}[b]{0.6\textwidth}
        \includegraphics[width = \textwidth]{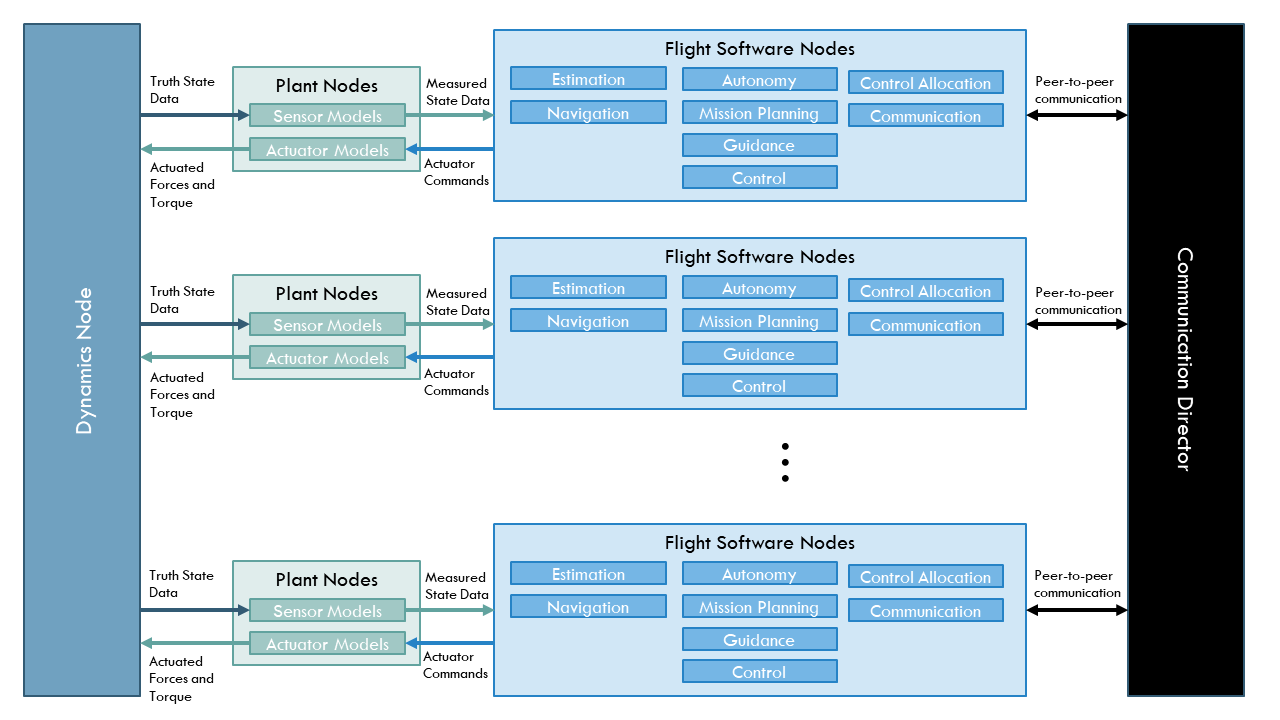}
        \caption{LINCS Simulation Environment}
        \label{fig:LINCS_Block_Diagram}
    \end{subfigure} 
    \begin{subfigure}[b]{0.43\textwidth}
        \includegraphics[width = \textwidth, trim = 4cm 0cm 5cm 0cm, clip]{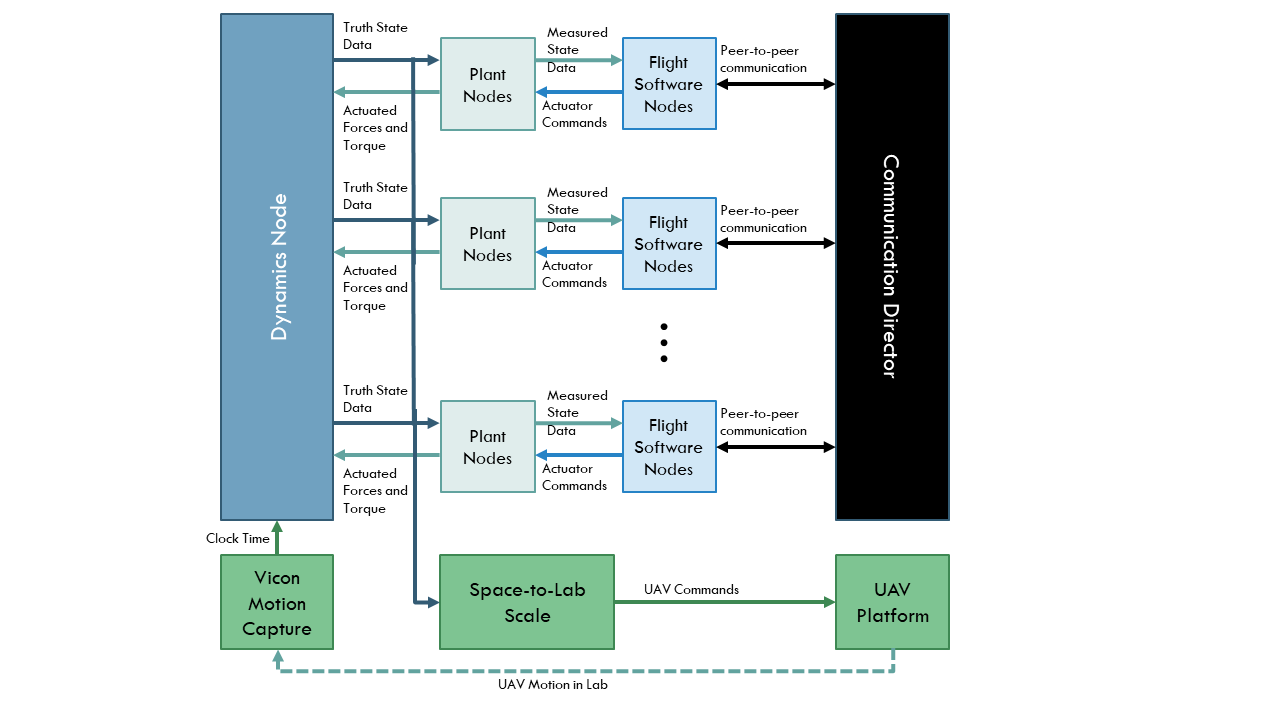}
        \caption{LINCS Emulation Environment}
        \label{fig:LINCS_Emulation}
    \end{subfigure}
    \caption{LINCS Environment block diagrams for information flow.}
    \label{fig:LINCS_block_diag}
\end{figure*}

LINCS aims to serve as a platform to perform verification, validation, and testing for autonomous space operations. % using a suite of analysis tools. 
As such, the lab is built to be accessible and has the ability to easily integrate various types of algorithms through the use of a modular ROS2 architecture.
Figure~\ref{fig:LINCS_block_diag} showcases the block diagrams for the software components of the laboratory. 
In particular, Figure~\ref{fig:LINCS_Block_Diagram}
shows the singular dynamics node responsible for dynamically handling variable numbers of
agents in any given experiment. 
The dynamics node drives the simulation by propagating the truth states of each satellite
in the simulation in an inertial frame, as well as, the location of celestial bodies in
the inertial frame. The truth states from the dynamics node are then passed to individual sensor plant nodes for each
satellite in the simulation as inertial states or relative states (i.e., resolved in Hill's frame). 
The sensor nodes then publish the simulated sensor readings to
the flight software nodes through the simulated satellite bus (i.e., ROS2 framework), allowing flight software nodes to query
sensor measurements as if they were running on a satellite. 
The flight software nodes include the autonomy, guidance,
navigation, control, or estimation algorithm being tested along with any other support nodes required to close the loop.
Support nodes might include estimation algorithms, low-level control policies, waypoint tracking algorithms, delta-v
conversions, or any other required nodes.
% 
% The ROS2 publisher and subscriber framework is not only how all the
% nodes in the full simulation communicate with each other, but also acts as the simulated satellite bus mimicking how the
% algorithm being tested would communicate with other applications when running on an actual satellite.
If required, the
flight software nodes can also utilize the singular communications director node to send custom messages back and
forth between the different agents to facilitate multi-agent planning and tasking.
Once the algorithm has generated
the desired control input, these are then converted into force and torque commands that are passed back into the plant nodes to be handled by actuator models, such as thruster models or CMG models. The resulting forces and torques actuated on the satellite given the
actuator commands are then sent back to the dynamics node, thus, completing the satellite simulation loop. 

Figure~\ref{fig:LINCS_Emulation} shows how the simulation environment couples with the emulation environment providing a valuable virtual to real pipeline to facilitate testing.
Notably, the motion capture system manages the real-time component of the simulation environment to ensure timing between real and virtual are synchronized. Additionally, there are added features that enable the measured lab vehicle states to feed back into the satellite flight software. This allows users of the lab to further understand how effective their algorithms are to real unmodelled effects invoked on the satellite systems, thus, providing real aspects to integration efforts on satellite systems.

\subsection{Integration in the Local Intelligent Network of Collaborative Satellites Laboratory}

The LINCS lab was used to demonstrate NNCSs trained using RL, and CBF-based RTA algorithms for spacecraft inspection \cite{dunlap2025demonstrating}. The control system was first tested in an open loop, where simulated data was passed to the NNCS and RTA to visualize the behavior of the controller on a real-world platform. The control system was then tested in a closed loop, with real world data passed to the NNCS and RTA. These simulations showed that the control system was robust to disturbances, where the agent was able to stay on task while adhering to safety constraints. % \todo{Nate add more description if necessary} - Nate: I think that covers things
Inspired by the LINCS lab, work was done to integrate the RL algorithms with Crazyflie drones via simulation testing \cite{de2024testing}, as a less expensive alternative testing path that may be integrated in university laboratories for cooperative research.

\begin{figure}
    \centering
    \includegraphics[width=\linewidth]{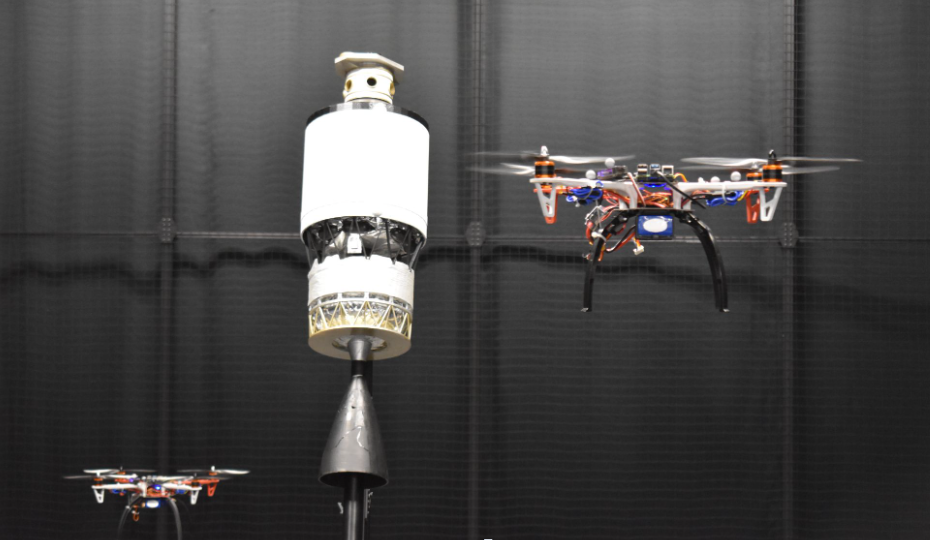}
    \caption{Two drones fly simulated spacecraft relative to a simulated resident space object inside the cage at the Local Intelligent Network of Collaborative Satellites Laboratory}
    \label{fig:LINCS}
\end{figure}

%\subsection{Integration of the Reinforcement Learning, Run Time Assurance, and Human-Artificial Intelligence Interface}
% \todo{Kyle and Nate - some description of the unique challenges in integration}

To demonstrate the HAI capabilities for RL and RTA, all these systems were integrated with the LINCS lab. The LINCS lab provides a standard interface for all groups, and allows testing the system on a real-world platform. For RL, the HAI allows users to change which NNCS is controlling an agent. For RTA, the HAI allows users to change safety constraint parameters, turn constraints on or off, and/or adjust the priority of each constraint. For each change, the HAI sends a message to the RL or RTA controller, which changes the system in real time. This capability allows operators to quickly adapt to changing mission objectives. Additionally, the HAI allows users to request data/updates about the control system at any time.

\subsection{Integration in the Spacecraft Performance Analytics and Computing Environment Research Laboratory}

One of the critical challenges to the deployment of autonomy technologies in space is developing algorithms that fit within the constrained processing and memory capacities of spacecraft processors. For spacecraft processors, there is typically a trade-off between how resilient the processor is to the high radiation environment of space and how much processing and memory is available. \textit{Radiation-hardened (rad-hard)} processors are the most robust in radiation environments, but tend to have the lowest processing capacity - often a decade or more behind terrestrial processors. \textit{Radiation-tolerant (rad-tolerant)} devices are not designed to last as long in a radiation environment and may produce more errors, but generally have more capable processors. Finally, for some short-duration missions in orbits below the Van Allen radiation belt formed by Earth's magnetic field, a variety of \textit{commercial-off-the-shelf (COTS)} processors may be used, which provide the greatest processing capability, but lack guarantees on resilience to radiation interference.

\begin{figure}
   \centering
   \includegraphics[width=\linewidth]{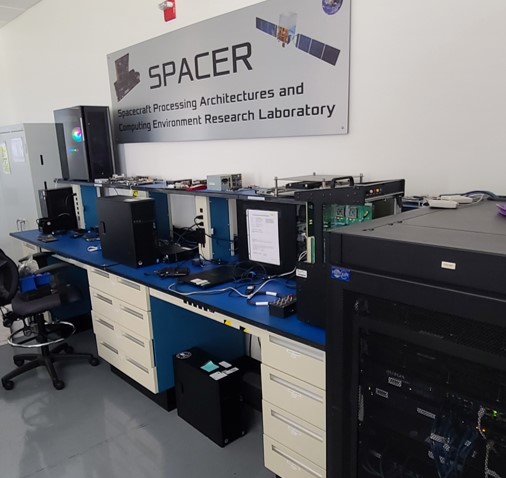}
   \caption{Processor testing in the Spacecraft Processing Architectures and Computing Environment Research (SPACER) Laboratory.}
   \label{fig:spacer}
\end{figure}

% \todo{The SPACER lab description} \nph{How's this short description? I can find more if needed}
% From https://www.kirtland.af.mil/Portals/52/documents/SPACERFactsheet.pdf?ver=2017-01-18-121826-480:
% "To leverage these capabilities for space, the Air Force Research Laboratory’s (AFRL) Space Vehicles Directorate (RV) has established an in-house computing architecture laboratory known as the Spacecraft Performance Analytics and Computing Environment Research (SPACER) testbed to characterize the utility of these diverse architectures for space applications.
% The SPACER testbed is equipped with commercial state-of-the-art and space grade processing platforms, which enables AFRL researchers to characterize the performance of sensitive application codes and lower level computational algorithms on relevant hardware. The development boards are networked together, when possible, to enable distributed computing applications across multiple platforms. The testbed has development workstations, hosting the development tools needed for programming the boards, in an enclave within the Defense Research and Engineering Network (DREN) to allow collaboration within AFRL.
The SPACER Laboratory was established in AFRL's Space Vehicles Directorate (RV) to characterize the utility of diverse computing architectures for space applications. To this end, the SPACER lab is equipped with a host of COTS, rad-tolerant, and rad-hard processors for the express purpose of testing and characterizing the performance of applications and algorithms on relevant hardware.

To validate that both the RL-trained NNCSs and the RTA algorithms, which use more computationally expensive optimization algorithms, function on processing- and memory-limited spacecraft hardware, the STARS team integrated both onto COTS and rad-tolerant processors in the SPACER lab. 
% 
% \todo{Nate and Kyle - add brief description of SPACER lab results /lessons learned \cite{dunlap2025space}}
% 
The results, detailed in \cite{dunlap2025space}, demonstrated that the trained NNCSs and a majority of RTA configurations designed for spacecraft inspection have a maximum observed execution time %\footnote{\begin{samepage} Maximum Observed Execution Time (MOET) is a metric used to approximate Worst-Case Execution Time (WCET) in systems without guaranteed scheduling. MOET and WCET are used to determine the minimum amount of time that should be allocated to execute the algorithm to ensure an answer is provided before it is needed. \end{samepage}}
well below 2 milliseconds on both the COTS and rad-tolerant processors. Therefore, these controllers can be used in a control system operating at a frequency up to 10 Hz, which is fast enough for the real-time operation of the spacecraft inspection system \cite{dunlap2025space}. The results also showed how a primary controller that produces safe actions reduces the amount of time needed for RTA to solve the optimization problem and therefore improves performance.
% Based on the maximum observed execution time, it was determined that an NNCS and most RTA configurations designed for spacecraft inspection can run at a minimum of 10 Hz on both the COTS and rad-tolerant processors, which is deemed to be fast enough for real time operation \cite{dunlap2025space}. It was also found that using a primary controller that produces safe actions reduces the amount of time needed for RTA to solve the optimization problem, and therefore improves performance.  \todo{Nate add more description if necessary}

\subsection{Safe Attitude Keep out Zones on GATech ASTROS}
The Georgia Institute of Technology ASTROS lab \cite{doi:10.2514/6.2009-1869} is a 5-degree-of-freedom platform for testing attitude and position GNC algorithms. This platform was used to test safe attitude RTA algorithms from \cite{10365665} and demonstrated that the design tools were able to keep the sensors out of the conical region keep-out zone. 

\begin{figure}
    \centering
    \includegraphics[width=1\linewidth]{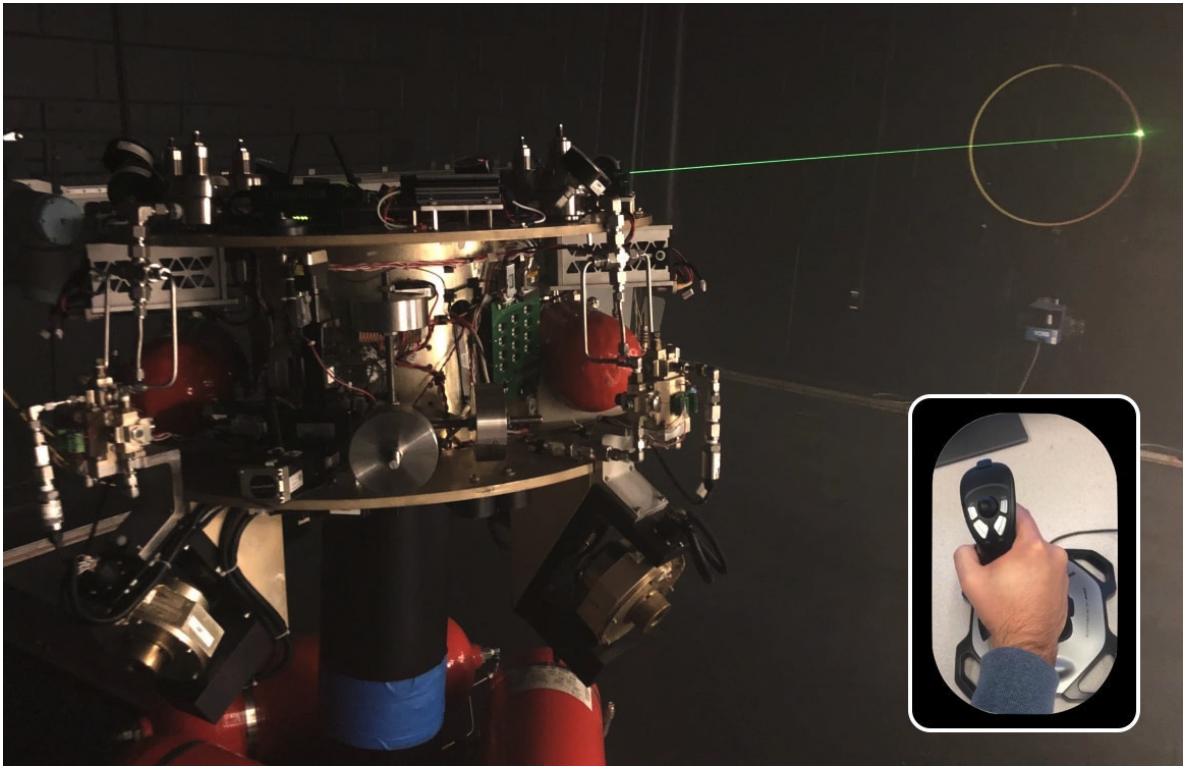}
    \caption{Experimental setup showcasing the attitude keep out algorithms from {\protect\cite{10365665}}. In this setup, the ring depicted the conical keep outzone as a cross-sectional area of the cone relative to the ASTROS platform. The green laser is the body-fixed direction instantiating the sensor foresight. The joystick is the "unverified" controller given by human operators.}
    \label{fig:RTA-attitude}
\end{figure}
%%%%%%%%%%%%%%%%%%%%%%%%%%%%%%%%%%%%%%
\section{Conclusion}
%%%%%%%%%%%%%%%%%%%%%%%%%%%%%%%%%%%%%%
The STARS program had the ambitious goal to develop and integrate a primary RL-based autonomous multi-vehicle controller, RTA to simultaneously assure multiple safety constraints, and HAI for spacecraft proximity operations, and evaluate the integrated system in a satellite dynamics emulation laboratory, and on spacecraft processors. Several RL-based implementations were developed that translate high level mission intent to decisions and control actions as deep in the guidance navigation and control architecture as force and moment commands.

Some of the most significant contributions of this work are in the RTA designs. While RTA is used here to bound the RL-based algorithm control actions to mission and safety constraints, it is designed modularly such that it can be implemented to enforce constraints on any primary controller, including scripted commands. In contrast to RTA designs that feature a specific reaction to prevent each possible constraint violation separately, the RTA ASIF controller finds solutions to many safety constraints simultaneously while being minimally invasive to the primary controller.  When the primary controller is safe, the RTA computation time is reduced. Comparisons of RL training with and without RTA revealed that some RTA constraints are useful for RL training as they can prevent irrelevant/undesirable states. Comparisons of continuous and discrete CBFs found that discrete CBFs can be used when RTA must be run at a low frequency, but results from space grade hardware show that continuous CBFs run fast enough for real-time operation.

Future work is needed to analyze the human-AI teaming interface to understand its effectiveness in facilitating directability, bounded autonomy, enhanced situation awareness, and calibrated trust in the autonomous agent's decision making at execution time. A planned future study with experienced spacecraft operators is expected to strongly influence future iterations of the interface. 

The RL-based multiagent autonomous spacecraft controllers, RTA safety algorithms, and human-AI teaming interface were all integrated into the LINCS lab for testing on higher fidelity simulation, as well as on physical vehicles that provided real-time feedback to evaluate the robustness of the solution in the presence of noisy state information. The combined system exceeded performance expectations. Additionally, tests on COTS and radiation-tolerant processors in the SPACER lab revealed that a train on the ground and deploy at the edge paradigm is feasible for NNCSs, and that the optimization-based ASIF RTAs can run sufficiently fast at the edge to be utilized in continuous control.

%\appendices{}              % note there is no {} to put a title. Each appendix has its own title
\section*{Appendix I: Acronyms}

\begin{table}[h!]
    \centering
    \label{tab:acronyms1}
    \begin{tabular}{p{1.2cm}P{6cm}}
        AEGIS & Autonomous Exploration for Gathering Increased Science\\
        AI & Artificial Intelligence\\
        %ANGELS & Automated Navigation and Guidance Experiment for Local Space \\
        %API & Application Programming Interface \\
        %ARL & Algorithm Readiness Level \\
        %ARPOD & Autonomous Rendezvous, Proximity Operations, and Docking\\
        ASIF & Active Set Invariance Filter \\
        ASTERIA & Arcsecond Space Telescope Enabling Research in Astrophysics \\
        ASTROS & Autonomous Spacecraft Testing of Robotic Operations in Space \\ 
        %C2BCM & Command Control Battle Management Communications\\
        CBF & Control Barrier Function \\
        CI/CD & Continuous Integration and Continuous Development\\
        COTS & Commercial-off-the-shelf\\
        %CRL& Commercialization Readiness Level\\
        %DAF & Department of the Air Force\\
        DARPA & Defense Advanced Research Projects Agency \\
        DART &Demonstration of Autonomous Rendezvous Technologies \\
        %DRL & Data Readiness Level\\
        %EAGLE & ESPA Augmented Geostationary Laboratory Experiment\\
        EDL & Entry, Descent and Landing\\
        %EELV &Evolved Expendable Launch Vehicle \\
        %ESA & European Space Agency\\
        %ESPA & EELV Secondary Payload Adaptor\\
        %FRP & Full Rate Production\\
        GNC & Guidance, Navigation and Control \\
        GPS & Global Positioning System\\
        HAI & Human-AI teaming Interface \\
        HIL & Hardware-in-the-loop \\
        HMT & Human-Machine Teaming \\ 
        %HRL & Human Readiness Level\\
        LINCS & Local Intelligent Network of Collaborative Satellites [Laboratory]\\
        %LVC & Live-Virtual-Constructive\\
        %LRIP & Low Rate Initial Production  \\
        ML & Machine Learning\\
        %MLRL & Machine Learning Readiness Level \\
        %MRL & Manufacturing Readiness Level\\
        MER & Mars Exploration Rover\\
        NASA & National Aeronautics and Space Administration\\
        NNCS & Neural Network Control System  \\
        PIL & Processor-in-the-loop \\
        %NRO & National Reconnaissance Office\\
        %P-LEO & Proliferated Low Earth Orbit\\
        %R\&D & Research \& Development\\
        %RAX & Remote Agent Experiment\\
        %RL & Readiness Level\\
        QP & Quadratic Program \\
        RL & Reinforcement Learning \\
        RTA & Run Time Assurance \\
        %RSGS & Robotic Servicing of Geosynchronous Satellites\\
        %SAE & Society of Automation Engineers \\
        %SDA & Space Domain Awareness \\
        %SSN & Space Surveillance Network\\
        SIL & Software-in-the-loop \\
        SPACER & Spacecraft Processing Architectures and Computing Environment Research (Laboratory) \\
        STAR  & Space Trusted Autonomy Readiness\\
        STARS & Safe Trusted Autonomy for Responsible Spacecraft \\
        %STM & Space Traffic Management \\
        %SWaP & Size, Weight and Power \\
        %TCL & Transition Commitment Level\\
        TRL & Technology Readiness Level\\
        TRN & Terrain Relative Navigation \\
        TrRL & Trust Readiness Level \\
        %UAS & Unmanned Aircraft System \\
        %USAF & United States Air Force \\
        USSF & United States Space Force \\
        V\&V & Verification and Validation \\
        %XSS-10 & Experimental Satellite System-10\\
        %XSS-11 & Experimental Satellite System-11\\
    \end{tabular}
\end{table}

%%%%%%%%%%%%%%%%%%%%%%%%%%%%%%%%%%%%%%%%%%%%%%%%%%%%%%%%%%%%%%%%%%%%%%%%%%%%%%%%%%%%%%%%%%%%%%%%%%%%%%
\acknowledgments
This research was sponsored by the Air Force Research Laboratory under the \textit{Seedlings for Disruptive Capabilities Program}. The views expressed are those of the authors and do not reflect the official guidance or position of the United States Government, the Department of Defense, or of the United States Air Force. Approved for public release; distribution unlimited. Case Numbers AFRL-2024-5123 and AFRL-2024-5659.

%%%%%%%%%%%%%%%%%%%%%%%%%%%%%%%%%%%%%%%%%%%%%%%%%%%%%%%%%%%%%%%%%%%%%%%%%%%%%%%%%%%%%%%%%%%%%%%%%

%%%%%%%%%%%%%%%%%%%%%%%%%%%%%%%%%%%%%%%%%%%%%%%%%%%%%%%%%%%%%%%%%%%%%%%%%%%%%%%%%%%%%%%%%%%%%%%%%%%%%%

\bibliographystyle{IEEEtran}
\bibliography{references}

%%%%%%%%%%%%%%%%%%%%%%%%%%%%%%%%%%%%%%%%%%%%%%%%%%%%%%%%%%%%%%%%%%%%%%%%%%%%%%%%%%%%%%%%%%%%%%%%%%%%%%
\thebiography
%% This biostyle allows you to insert your photo size 1in X 1.25in
\begin{biographywithpic}{Kerianne L. Hobbs}{Figures/Bios/HobbsKerianne8x10}
is the Safe Autonomy and Space Lead on the Autonomy Capability Team (ACT3) at the Air Force Research Laboratory. There she investigates methods to evaluate safety, trust, ethics, and performance of autonomous systems. Her previous experience includes work in automatic collision avoidance technologies for F-16s and autonomy verification and validation research. Dr. Hobbs’s research has resulted in authorship of over 60 peer reviewed publications, conference papers, technical magazine articles (IEEE Control Systems Magazine, AIAA Aerospace America) and strategy documents, as well as over 60 invited or conference presentations. Dr. Hobbs was selected for the 2024 AIAA Associate Fellow Class, 2020 AFCEA 40 Under 40 award, and was a member of the team that won the 2018 Collier Trophy (Automatic Ground Collision Avoidance System Team). Dr. Hobbs has a BS in Aerospace Engineering from Embry-Riddle Aeronautical University, an MS in Astronautical Engineering from the Air Force Institute of Technology, and a Ph.D. in Aerospace Engineering from the Georgia Institute of Technology.
\end{biographywithpic}

\begin{biographywithpic}{Joseph B. Lyons}{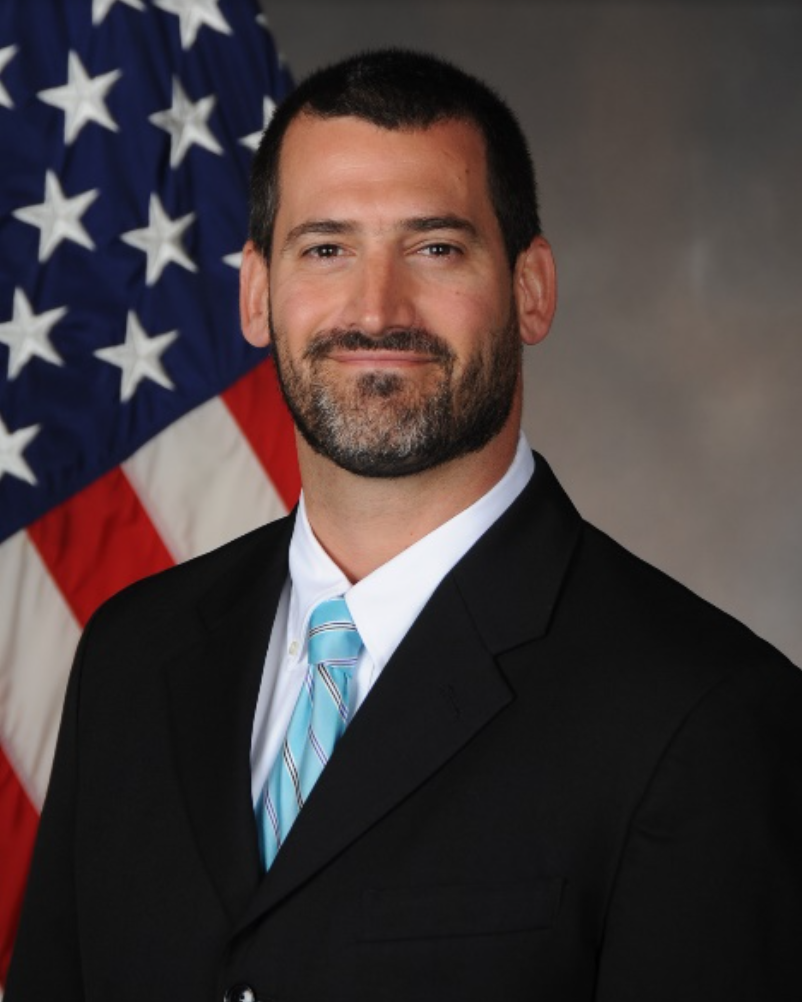} is a member of the scientific and professional cadre of senior executives, is the Senior Scientist for Human-Machine Teaming, 711th Human Performance Wing, Human Effectiveness Directorate, Air Force Research Laboratory, Wright-Patterson AFB, Ohio.  He serves as the principal scientific authority and independent researcher in the research, development, adaptation, and application of Human-Machine Teaming.  Dr. Lyons began his career with the Air Force in 2005 in the Human Effectiveness Directorate, Wright-Patterson AFB, Ohio. Dr. Lyons has served as a thought leader for the DoD in the areas of trust in autonomy and Human-Machine Teaming. Dr. Lyons has published over 100 technical publications including 64 journal articles in outlets focused on human factors, human-machine interaction, applied psychology, robotics, and organizational behavior. Dr. Lyons also served as Co-Editor for the 2020 book, Trust in Human-Robot Interaction. Dr. Lyons is an AFRL Fellow, a Fellow of the American Psychological Association, and a Fellow of the Society for Military Psychologists. Prior to assuming his current position, Dr. Lyons served as a Program Officer for the Air Force Office of Scientific Research and was a Principal Research Psychologist within the Human Effectiveness Directorate. 
\end{biographywithpic}

\begin{biographywithpic}{Sean Phillips}{Figures/Bios/Phillips_img} is the Technology Advisor and Senior Mechanical Engineer for the Space Control Branch. He leads the Local Intelligent Networked Collaborative Satellites (LINCS) Lab and the Distributed Resilient Multi-Satellite Autonomy Program at the Air Force Research Laboratory. He is a Research Assistant Professor (LAT) at the University of New Mexico in Albuquerque, NM. He received his Ph.D in the Department of Computer Engineering at the University of California – Santa Cruz in 2018. He received his B.S. and M.S. in Mechanical Engineering from the University of Arizona in 2011 and 2013, respectively. He received the AFRL Early Career Award in 2024. Dr. Phillips's current research interests consist of robust GNCA algorithm development, hybrid system modeling and analysis, satellites systems, communication/information networks, and distributed systems under adverse conditions. 
\end{biographywithpic}

\begin{biographywithpic}{Michelle Simon}{Figures/Bios/MichelleSimon} is the technical lead for the Autonomous Operations Resilience for Tactile Action Program. She has been with the Air Force Research Laboratory since 2016 and has over 10 years in the aerospace industry. She currently works in the Space Vehicles Directorate in Albuquerque, NM, where her primary area of research is autonomous systems. She received a B.S. in Computer Programming from Pacific Lutheran University in Tacoma, WA in 2010 and an M.S. in Electrical Engineering from University of Alaska in Fairbanks, AK in 2016.
\end{biographywithpic}

\begin{biographywithpic}{Kyle Dunlap}{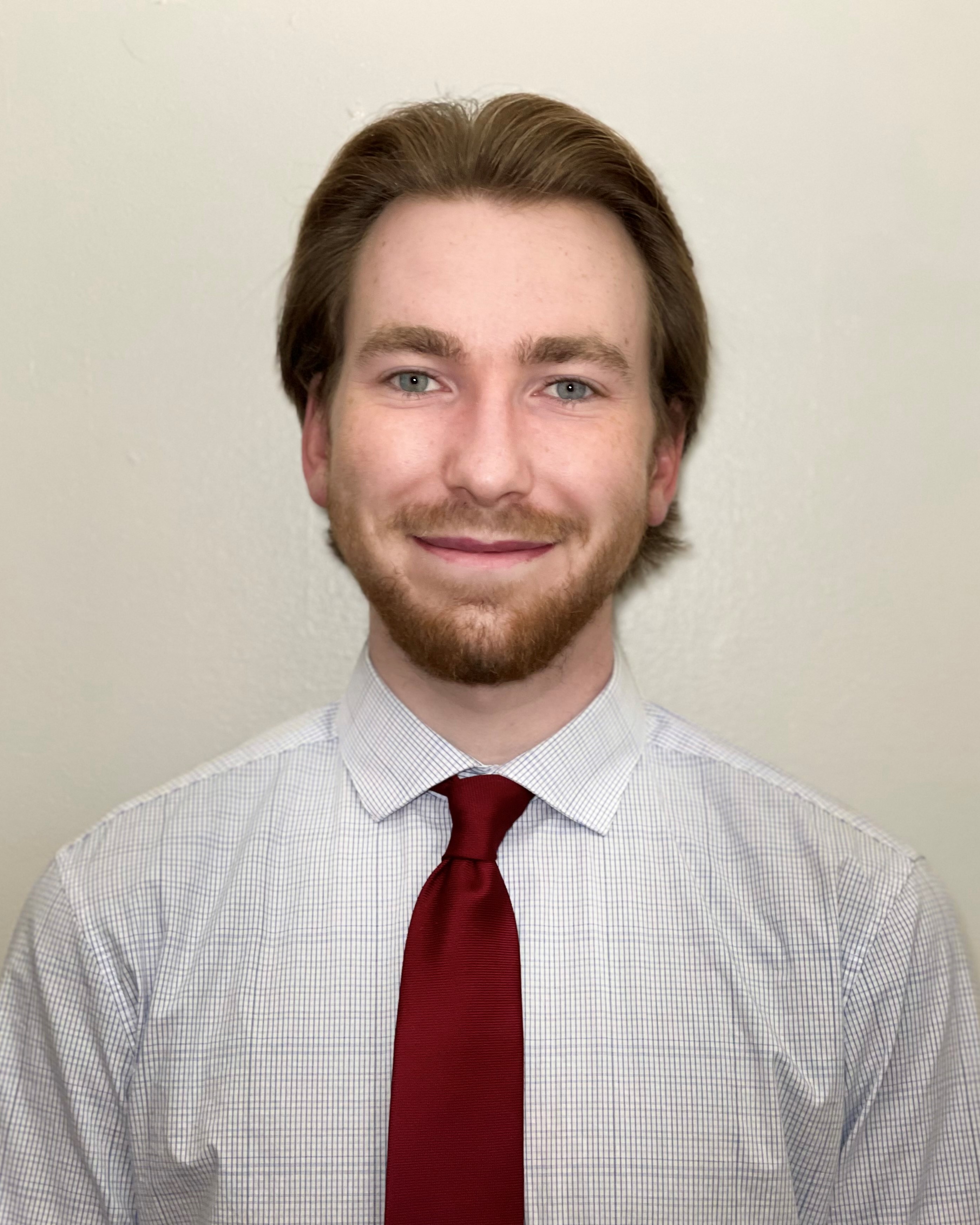}
is the Run Time Assurance Lead on the Safe Autonomy Team at the Air Force Research Laboratory's Autonomy Capability Team (ACT3). There he investigates real time safety assurance methods for intelligent aerospace control systems. His previous experience includes developing a universal framework for Run Time Assurance and comparing different RTA approaches during Reinforcement Learning training. He received his BS and MS in Aerospace Engineering from the University of Cincinnati.
\end{biographywithpic}

\begin{biographywithpic}{Nathaniel Hamilton}{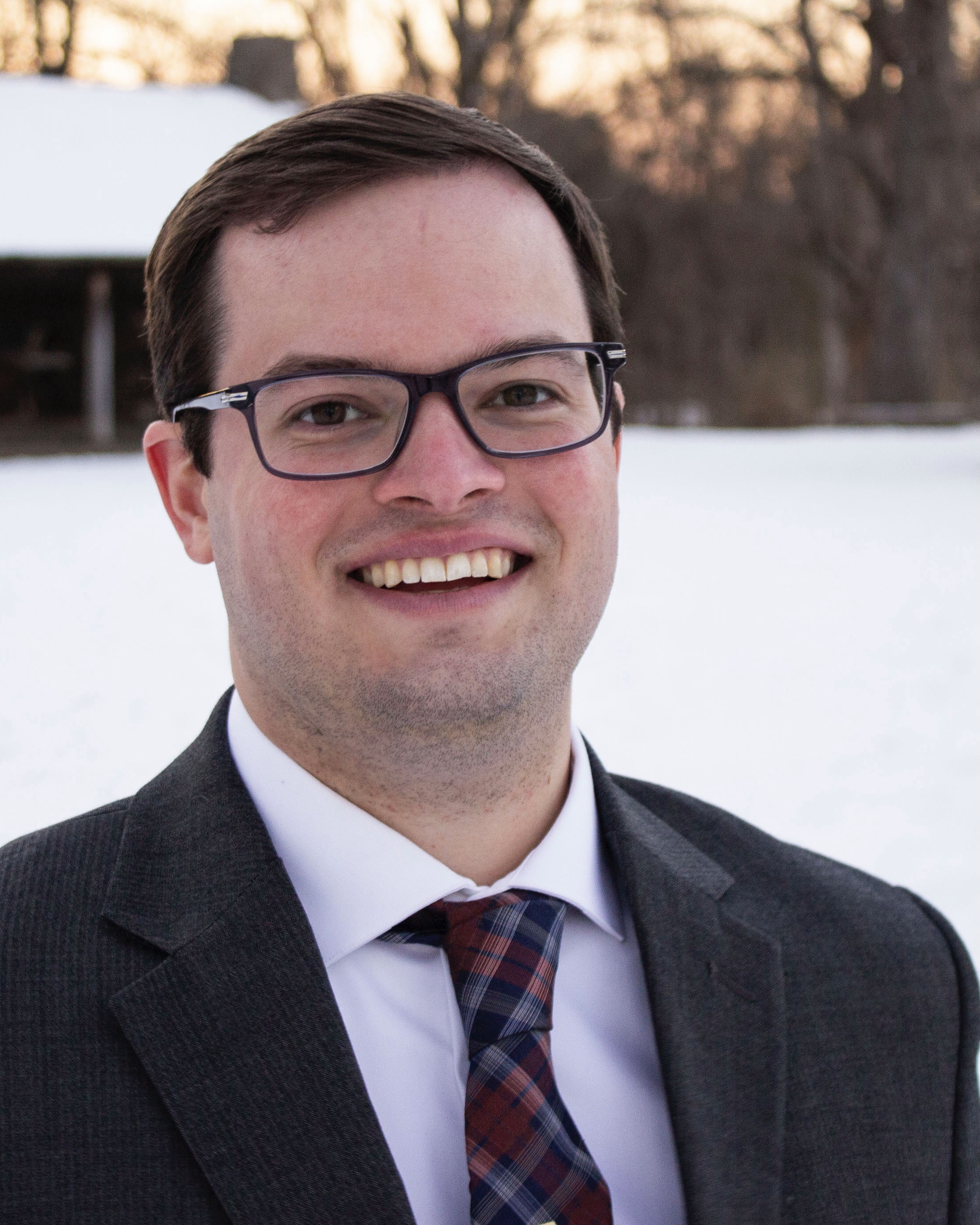}
% is the Safe Reinforcement Learning Lead on 
if an AI Scientist at Parallax Advanced Research supporting the 
Safe Autonomy Team at the Air Force Research Laboratory's Autonomy Capability Team (ACT3). There he investigates Safe Reinforcement Learning (SafeRL) approaches and how we can better integrate safety into the learning process to enable safe, trusted, and certifiable autonomous and learning-enabled controllers for aircraft and spacecraft applications. His previous experience includes studying how Run Time Assurance (RTA) impacts the learning and performance of SafeRL agents, and work in simulation to real-world (sim2real) transfer for learning-enabled controllers. In 2019, Dr. Hamilton was awarded the National Defense Science and Engineering Graduate (NDSEG) Fellowship. Dr. Hamilton has a BS in Electrical and Computer Engineering from Lipscomb University, and an MS and Ph.D. in Electrical Engineering from Vanderbilt University.
\end{biographywithpic}

\begin{biographywithpic}{Joshua Aurand}{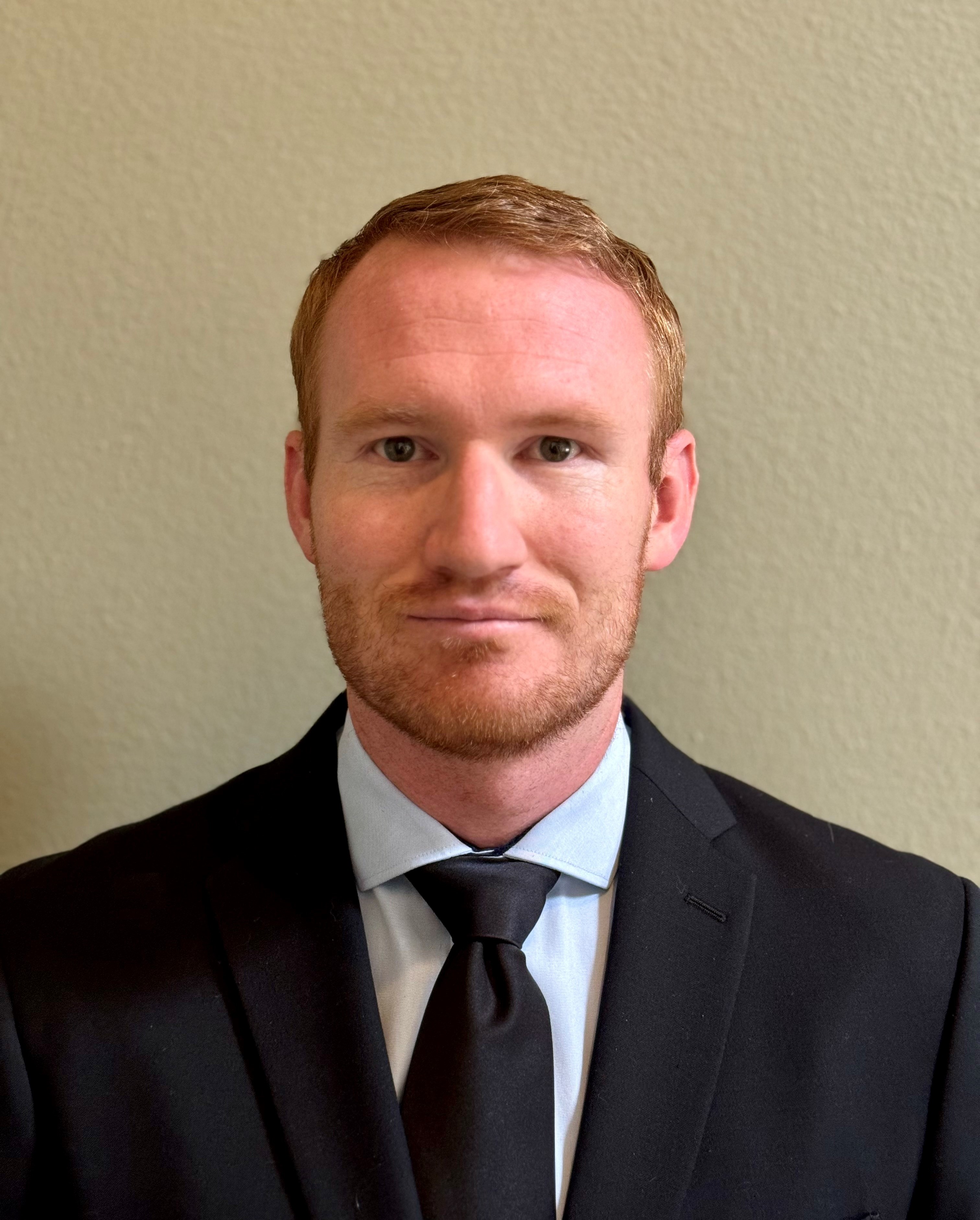}
completed his PhD in Applied Mathematics at University of Colorado at Boulder in 2020 and completed a  Postdoctoral Research appointment at Dublin City University in 2022. His research focus was on the impact of market shocks on optimal control strategies for heterogeneous sets of agents. Currently he is an Artificial Intelligence and Machine Learning Team Lead at Verus Research in Albuquerque, NM. His current research interests include developing resilient multiagent command and communication architectures for rendezvous and proximity operations. He also leads efforts to apply game theory and reinforcement learning to multi-satellite tasking.
\end{biographywithpic}

\begin{biographywithpic}{Jared Culbertson}{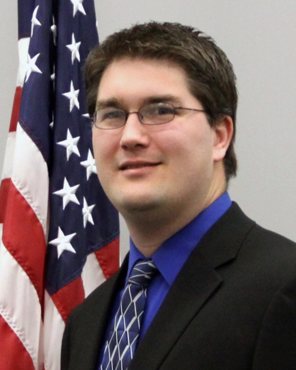} is a senior research mathematician on the Autonomy Capability Team (ACT3) at the Air Force Research Laboratory. He started at AFRL in 2010 in the Sensors Directorate prior to having a founding role in the launch of ACT3. His research primarily deals with fundamental aspects of representational structures, currently focused on robust behavior acquisition, diversity, and composition in reinforcement learning problems. Dr. Culbertson has a B.S. in mathematics from Harding University, as well as M.S. and Ph.D. degrees in mathematics from Louisiana State University.
    
\end{biographywithpic}

\begin{biographywithpic}{Zachary I. Bell}{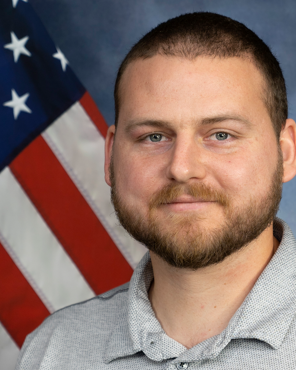} Zachary I. Bell received his Ph.D. from the University of Florida in 2019 and is a researcher for the Air Force Research Lab. His research focuses on cooperative guidance and control, computer vision, adaptive control, and reinforcement learning. 
\end{biographywithpic}
\vspace{3pt}
\begin{biographywithpic}{Taleri Hammack}{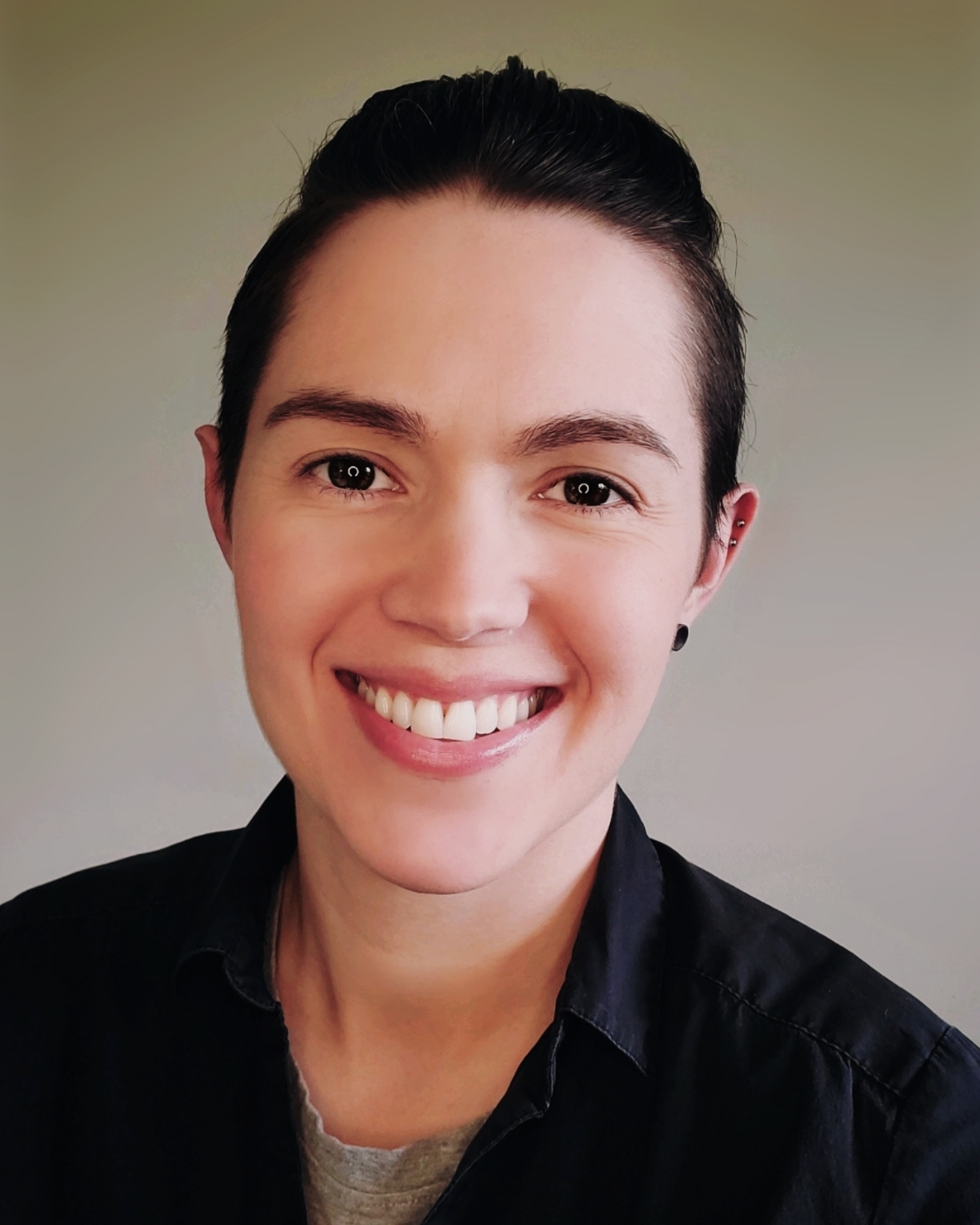}  is the lead for the Human-AI teaming Interface (HAI) design effort under the STARS program. There they work closely with space operators and autonomy developers to ensure autonomous solutions are understood and adaptable to an operator’s changing goals and priorities for the mission. Dr. Hammack has over 10 years of experience in cognitive systems engineering focused on human-autonomy teaming spanning a variety of DoD areas including space operations, Army air assault missions, and joint all-domain command and control systems. Dr. Hammack earned their Ph.D in Human Factors Psychology at Wright State University in 2022, and their M.S. in 2015, and a B.S. in Psychology from the University of Idaho in 2012. \end{biographywithpic}

\begin{biographywithpic}{Dorothy Ayres}{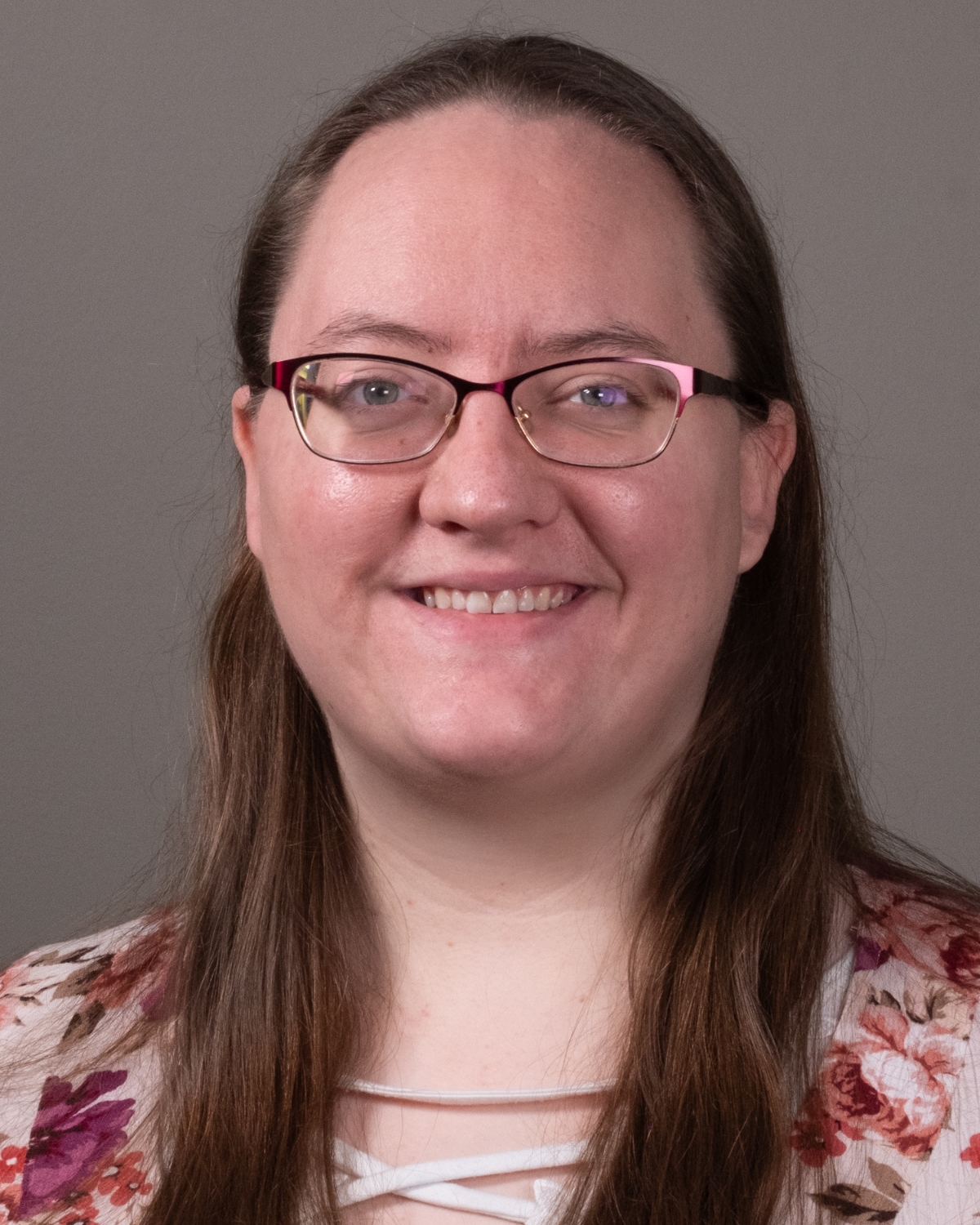} Dorothy Ayres is a member of the team working on the Human-AI teaming interface design for the STARS program. She is a human factors scientist working for DCS Corporation, where she works in HMI design, focusing on human factors principles to design interfaces to facilitate communication and situational awareness for operators. Dorothy has been with DCS in human factors interface design for over three years. She received an M.S. in human factors psychology from Wright State University in 2021, and a B.S in Psychology from Wright State University in 2018.\end{biographywithpic}

\begin{biographywithpic}{Rizza Lim}{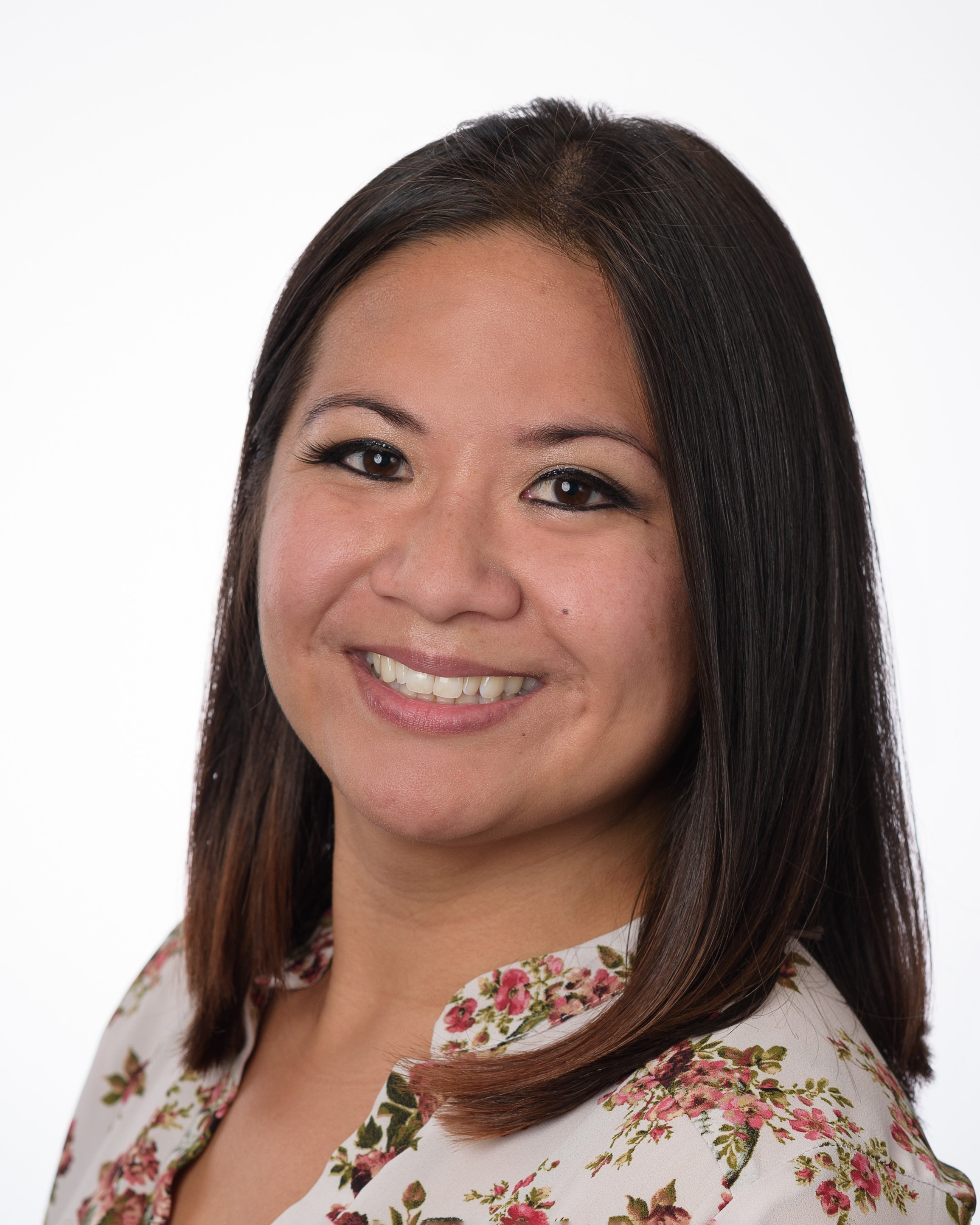} Rizza Lim is a UI/UX Designer within the Human-Autonomy Systems Department at DCS Corporation. There she leads the Design Strategy using user-centered design processes to solve user problems across different DOD products. Her experience includes user research, interaction and motion design, rapid prototyping, and workshop facilitation. She collaborated and designed products for the Shadow Operation Center (ShoC-N) at Nellis Air Force Base, Defense Advanced Research Projects Agency (DARPA), the Air Force's Advanced Battle Management System (ABMS), and the Air Force’s Digital Directorate known as Kessel Run. She served in the U.S. Navy as an IT from 2008 to 2016. She received her B.S. in Computer Information Systems from Park University in 2015 and her M.B.A from Walden University in 2018.\end{biographywithpic}

\begin{biographywithpic}{Zachary S. Lippay}{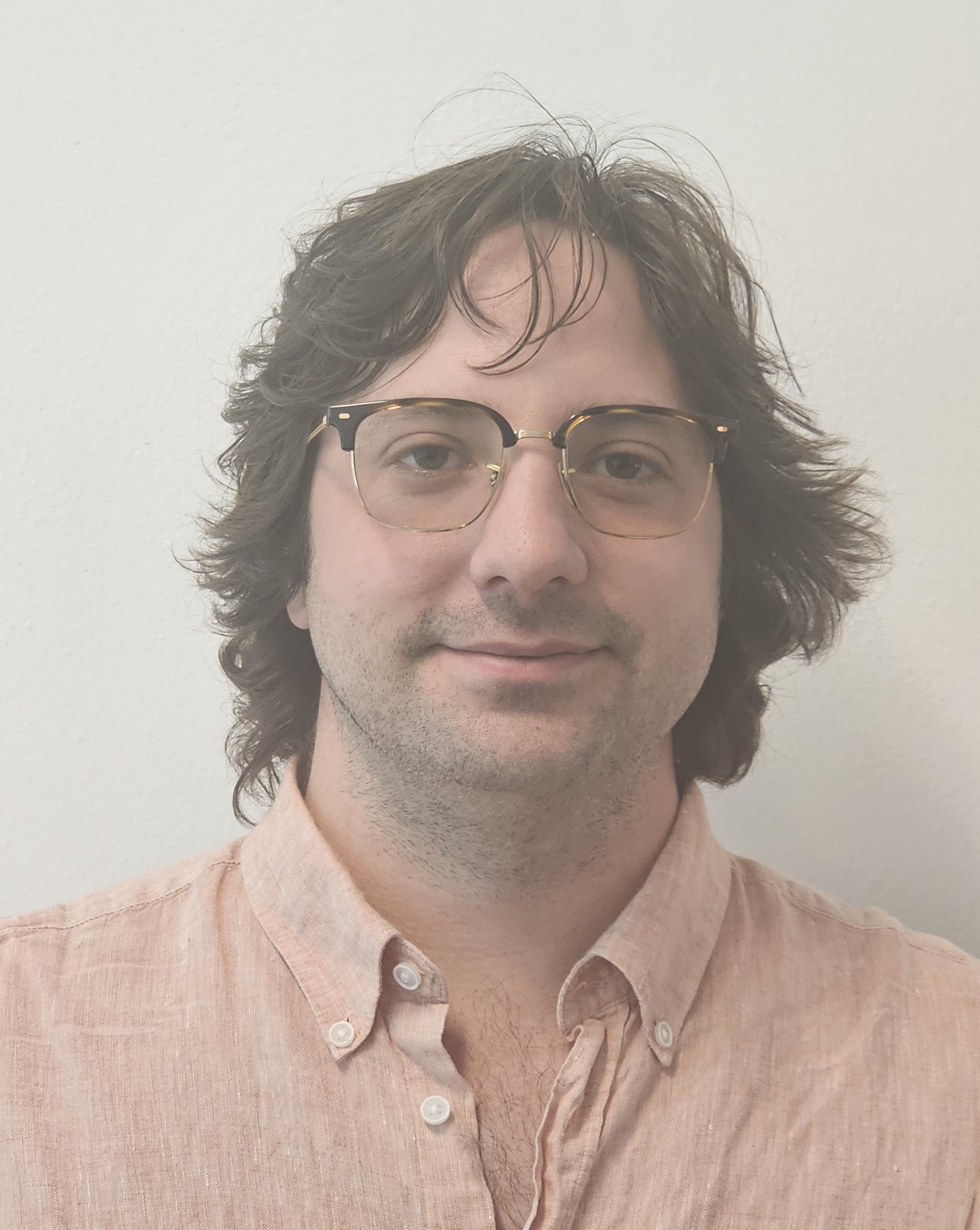} received his B.S. degree in Mechanical Engineering from the University of Kentucky in 2016. He received the Ph.D. degree in 2022 from the department of Mechanical Engineering at the University of Kentucky with a focus on cooperative and safe multiagent systems. He continued his work as a Research Controls Engineer at Verus Research in 2022 in the Space \& Autonomy portfolio where the team develops algorithms for autonomous satellite operations, as well as develops testing, evaluation, verification, and validation techniques for satellite autonomy. Currently he serves as the Dynamics and Controls Team Lead in the Space \& Autonomy portfolio at Verus Research in Albuquerque, NM. \end{biographywithpic}

\end{document}